\documentclass[twocolumn]{aastex63}

\usepackage{rotating}

\shorttitle{Double Detonation SNe Ia}
\shortauthors{Boos et al.}
\graphicspath{{./}{figures/}}

\begin{document}

\title{Multi-Dimensional Parameter Study of Double Detonation Type Ia Supernovae Originating from Thin-Helium-Shell White Dwarfs}

\author[0000-0002-1184-0692]{Samuel J. Boos}
\affiliation{Department of Physics \& Astronomy, University of Alabama, Tuscaloosa, AL, USA}

\author[0000-0002-9538-5948]{Dean M. Townsley}
\affiliation{Department of Physics \& Astronomy, University of Alabama, Tuscaloosa, AL, USA}

\author[0000-0002-9632-6106]{Ken J. Shen}
\affiliation{Department of Astronomy and Theoretical Astrophysics Center,
        University of California, Berkeley, CA, USA}

\author{Spencer Caldwell}
\affiliation{Department of Physics \& Astronomy, University of Alabama, Tuscaloosa, AL, USA}

\author[0000-0003-1278-2495]{Broxton J. Miles}
\affiliation{Applied Research Associates, Inc. Raleigh, NC, USA}










\begin{abstract}

Despite the importance of Type Ia supernovae (SNe Ia) throughout astronomy, the precise progenitor systems and explosion mechanisms that drive SNe Ia are still unknown. An explosion scenario that has gained traction recently is the double detonation in which an accreted shell of He detonates and triggers a secondary detonation in the underlying white dwarf. Our research presents a number of high resolution, multi-dimensional, full star simulations of thin-He-shell, sub-Chandrasekhar-mass white dwarf progenitors that undergo a double detonation. This suite of thin-shell progenitors incorporates He shells that are thinner than those in previous multi-dimensional studies. We confirm the viability of the double detonation across a range of He shell parameter space as well as present bulk yields and ejecta profiles for each progenitor. The yields obtained are generally consistent with previous works and indicate the likelihood of producing observables that resemble SNe Ia. The dimensionality of our simulations allow us to examine features of the double detonation more closely, including the details of the off-center secondary ignition and asymmetric ejecta. We find considerable differences in the high-velocity extent of post-detonation products across different lines of sight. The data from this work will be used to generate predicted observables and may further support the viability of the double detonation scenario as a SNe Ia channel as well as show how properties of the progenitor or viewing angle may influence trends in observable characteristics.

\end{abstract}



\section{Introduction}

Type Ia supernovae (SNe Ia) are some of the most luminous and consequential events in astronomy, but the mystery regarding their exact origin remains. 
While the source behind their extreme luminosity and characteristic light curve has been known for decades \citep{Pankey_1962,Colgate_etal_1969}, there has yet to be a complete and precise explosion mechanism and progenitor system model that can fully explain the origin and nature of SNe Ia.
This can be disconcerting among the numerous and diverse areas of research that rely on observations and trends of SNe Ia.
A more accurate depiction of the origin of SNe Ia would serve to improve their contributions to many areas of research, including nucleosynthesis, galaxy evolution, and cosmology.

SNe Ia distinguish themselves from other supernovae types in the lack of H and He lines in their spectra, indicating a C-O white dwarf progenitor \citep[see][for a review]{Maoz_14}.
Additionally, SNe Ia have distinctive Si, Ca, and Fe spectroscopic features \citep{Parrent_etal_2014}.
A notable trend in the light curves of SNe Ia is the Phillips relation in which more luminous SNe Ia decline more slowly from peak brightness \citep{Phillips_93}.
This unique characteristic allows the distance modulus of a SN Ia to be calculated from its decline rate and apparent peak magnitude.
The general trend embodied in the Phillips relation arises from the tendency of Fe-group elements, particularly radioactive $^{56}$Ni, to provide both the energy source and the principal component of opacity in the remnant ejecta.
Thus a single parameter sets the brightness, the diffusion time \citep{Hoeflich_etal_1996,Pinto_Eastman_2001}, and the time for ionization transitions effecting color evolution \citep{Kasen_Woosley_2007}.
However, such a deterministic trend implies an underlying systematic variation of the ejecta profile with $^{56}$Ni mass that itself requires a deeper explanation from the explosion mechanism \citep{Woosley_Kasen_etal_2007}.
A comparable relation between the decline rate and the intrinsic color of the SNe Ia is similarly critical in correcting for extinction \citep{Jha_Riess_Kirshner_2007}.
As the full mechanics that make SNe Ia into this nearly single-parameter family are unclear, further understanding of them is beneficial to the role of SNe Ia as a rung in the cosmic distance ladder and in the determination of the Hubble constant \citep{Riess_etal_2016,Riess_etal_2019}, as well as evidence for the accelerating expansion of the universe \citep{Riess_1998,Perlmutter_1999,Betoule_etal_2014,Scolnic_etal_2018}.
Reproducing the Phillips relation, among other SN Ia traits, via simulation would serve to help determine the viability of the particular candidate model as well as reducing uncertainties in the calculation of SN Ia distances.

While the precise model is still a mystery, there is strong consensus that SNe Ia are the result of thermonuclear explosions in C-O white dwarfs (WDs) due to estimated energetic outputs and the aforementioned spectral features \citep{Maoz_14,Seitenzahl_Townsley_2017}.
Theories vary in both progenitor system and explosion mechanism.
A promising explosion scenario, which we will discuss here, is the double detonation. 
In the double detonation, an accreted layer of He at the surface of a C-O WD ignites and undergoes a thermonuclear runaway.
The secondary explosion in the core may be triggered directly at the core-shell interface or when the inward propagating shock from the initial explosion converges at some location within the core
(\citealt{Livne_Glasner_1991}; \citealt{Woosley_94}; \citealt{Fink_2007}; \citealt{Fink_10}; \citealt{Shen_2014a}; see \citealt{Townsley_2019a} for a historical overview).
Unlike a model that requires a primary core ignition, in this scenario WDs with masses below the Chandrasekhar mass can lead to SNe Ia, increasing the range of theoretically viable progenitor systems.
Ideally, the WD mass is the single parameter, so that the total $^{56}$Ni ejected and the overall ejecta profile vary together to give, roughly, a single sequence.

A new, compelling motivation for double detonation is the dynamically driven double degenerate double detonation (D$^6$) model \citep{Shen_2018b}.
In the D$^6$ model, the shell ignition is brought on by mass transfer from a degenerate companion \citep{Guillochon_etal_2010}.
As this scenario would occur in a tight binary, a supernova in the primary WD would unbind the binary system and eject the companion at a speed close to its pre-explosion orbital velocity. 
\cite{Shen_2018b} identified three possible candidates for these ``runaway'' stars in the Gaia astrometric survey, with one being loosely traced back to a supernova remnant.

Much research on double detonations has examined relatively thick He layers, which produce results that did not match observations of normal SNe Ia very well
\citep{Woosley_94,Hoeflich_1996,Nugent_etal_1997}. 
Also, some work strongly suggested that a thin enough He shell would produce a spectroscopically normal SN Ia \citep{Kromer_etal_2010,woosley_2011,Townsley_etal_2012,Moore_etal_2013}.
It was then found that the inclusion of key elements, particularly nitrogen, in the shell abundances and nuclear reaction network eased the constraints on the shell masses necessary to produce and host a detonation \citep{Shen_2014b}.
\cite{Townsley_2019a} explored the double detonation scenario by conducting a single, full-star simulation of a modestly enriched, thin-shell progenitor and found that these cases can indeed produce spectroscopically normal SNe Ia.
We follow up on that work here by exploring a broader set of progenitors.

In this work, we greatly extend the parameter space of the thin-shell SN Ia candidate from \cite{Townsley_2019a}. 
We simulate double detonations in progenitors with a wider range of masses and He shell thicknesses and abundances, generating isotopic yields and 2-dimensional ejecta profiles to be used in the generation of predicted observables.
The shell masses used in our thin-shell progenitors are lower than in previous multi-dimensional studies, enabled by our use of an expanded nuclear network.
We discuss the details of our simulations in Section \ref{sec:methods}, including the progenitors, starting conditions, and methods used.
Results of the simulations such as yields and profiles, as well as remarkable details of the explosion process and comparison to other work, are discussed in Section \ref{sec:results}.
Conclusions, in addition to some discussion on consequential results, are highlighted in Section \ref{sec:conclusions}.

\section{Methods}
\label{sec:methods}

This study builds on our previous work simulating supernovae and uses the astrophysical reactive fluid dynamics software \texttt{FLASH} and the nuclear reaction networks from \texttt{MESA}.
In this section we detail both the chosen starting point of our simulation, that is the progenitor WD and He ignition, as well as the methods used to simulate the explosion and compute yields and profiles.
Many of these are as used in previous work, but we also note some changes new to this work, specifically the improved treatment of the domain-filling ``fluff.''

\subsection{Progenitors}

The parameters of the eleven C-O WD progenitors used in this study are shown in Table \ref{table:parameters}.
Progenitors vary in total, core, and shell mass as well as He shell geometric thickness as determined by the temperature at the base of the shell.
They can be categorized into two shell base layer density regimes: thin (2 and $3\times 10^5$ g~cm$^{-3}$) and thick (6 and $14\times 10^5$ g~cm$^{-3}$).
The nine thin cases are the thin-shell candidate progenitors of SNe Ia at the focus of this study. 
The two thick shell cases provide a useful comparison to the many other simulation studies that examined thick shell detonations (e.g. \cite{Fink_2007,Kromer_etal_2010,Polin_2019}).

\begin{table}
\centering
\caption{Progenitors}
\begin{tabular}{ccccc}
M$_{tot}$ & M$_{shell}$ & $\rho_{base}$  & $T_{base}$  & Expansion Time \\ 
($M_\odot$) & ($M_\odot$) & ($10^5$ g cm$^{-3}$) & ($10^8$ K) & (s) \\ \hline \hline
0.85 & 0.033 & 2 & 3.0 & 21.4  \\
0.90 & 0.025 & 2 & 3.0 & 70.6 \\
1.00 & 0.016 & 2 & 3.5 & 97.1  \\
1.10 & 0.008 & 2 & 3.5 & 97.2 \\ \hline
1.02 & 0.021 & 2 & 5.0 & 31.3  \\ 
1.02 & 0.021\footnote{Decreased $^{16}$O} & 2 & 5.0 & 26.9 \\ \hline
0.85 & 0.049 & 3 & 3.0 & 97.0 \\ 
1.00 & 0.021 & 3 & 3.0 & 97.2 \\ 
1.10 & 0.011 & 3 & 3.0 & 97.6 \\ \hline 
1.00 & 0.042 & 6 & 2.7 & 97.8 \\
1.00 & 0.100 & 14 & 2.3 & 98.5 \\ 
\end{tabular}
\label{table:parameters}
\end{table}

The composition of all but one progenitor very closely matches that of the modestly enriched shell model from \citet{Townsley_2019a}.
By mass, the core composition is 0.4 $^{12}$C, 0.587 $^{16}$O, and 0.013 $^{22}$Ne.
The shell composition is 0.891 $^{4}$He, 0.05 $^{12}$C, 0.009 $^{14}$N, and 0.05 $^{16}$O.
This moderate enrichment enhances the ignition in the shell \citep{Shen_2014b}.
One thin-shell progenitor uses an alternative shell composition with reduced $^{16}$O of 0.015, with the difference made up in the He.
This may be more representative of the enrichment based on modest mixing between the He layer and the outer part of the core, which is made up of the C-rich ashes of the late burning stages on the asymptotic giant branch.
The progenitors are constructed to be hydrostatic with the same equation of state used by \texttt{FLASH} and thus necessitate no settling when initialized in \texttt{FLASH}.

\begin{figure}
    \centering
    \includegraphics{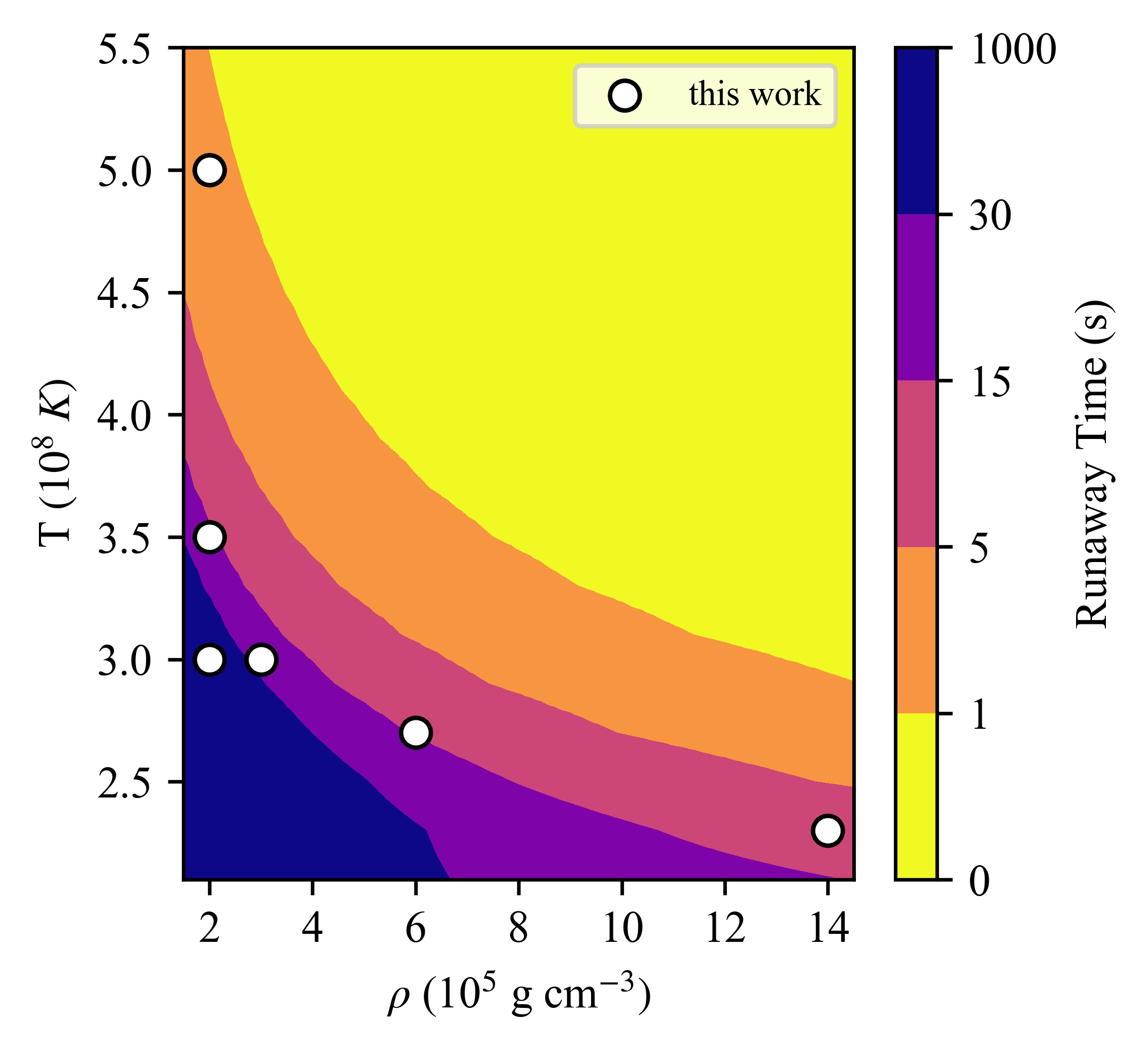}
    \caption{Runaway timescale contours for He shell material.
    The timescale was taken as the temperature runaway time in constant density self-heating simulations.
    Circles indicate the He shell base parameters of the progenitors used in this work.
    Some progenitors have the same shell base density and temperature, but differing core and shell masses or compositions, and overlap on this plot.
    }
    \label{fig:timescale}
\end{figure}

The core of each progenitor has a uniform temperature of $3\times 10^7$ K.
The He layer has an adiabatic temperature profile intended to mock up what may be present due to mixing induced by a dynamical mass transfer stream from the secondary \citep{Guillochon_etal_2010}.
The choice of base layer temperature is based on the local temperature runaway timescale determined from constant-volume self-heating calculations of material with composition, density, and temperature corresponding to the shell base layer (see Figure \ref{fig:timescale}).
For each shell base density, in most cases, the shell base temperature is chosen such that the corresponding local runaway timescale is tens of seconds.
This is greater than the time it takes the He denotation to propagate around the star, about 2 seconds, so that ignition only occurs at our chosen site.
Two cases have shorter runaway times, comparable to the propagation time around the star.
These are the case with a lower oxygen abundance and a standard composition case both with masses of  $1.02 \, M_\odot$, which have a higher base temperature, $5\times 10^8$ K.
This leads to a geometrically thicker and therefore more massive He shell for a given core mass and base density.
The question of the appropriate temperature state of the shell and further exploration of how the ignition takes place are left to separate work.

\subsection{Nuclear reaction network and limiter}
\label{sec:network}

We utilize the same 55-species nuclear reaction network in our simulations as \citet{Townsley_2019a}, designed to obtain accurate energy release for high-temperature He- and C-O-burning.
The nuclear reactions are integrated during each step in each cell using the constant temperature burning functionality provided by the \texttt{MESA/net} module version 9793.

Unlike in our previous work (\citealt{Townsley_2019a}, see Section \ref{subsec:t19_comp} for a comparison of yields) we use a burning limiter \citep{kushnir_head-collisions_2013} similar to that discussed by \citet{Shen_2018a} but with some modifications to make the energy release rate independent of the timestep and to be more efficient.
The burning integration is limited so that the energy change per mass in a single timestep $|\Delta \mathcal E|$ is no larger than $c_VT |\Delta\ln T|_{\rm max}\Delta t_{\rm hydro}/\Delta t_{\rm sound}$.
Here $c_V$ is the mass-specific heat, $|\Delta \ln T|_{\rm max}$ is chosen to be 0.1, $\Delta t_{\rm hydro}$ is the hydrodynamic timestep, and $\Delta t_{\rm sound}$ is the sound crossing time of the smallest cell in the simulation.
As a result, when the limiter is active, the energy release rate, $|\Delta \mathcal E/\Delta t_{\rm hydro}|$ only depends on the value chosen for the limit and the cell size, not the CFL or hydro timestep.
Since $\Delta t_{\rm hydro}$ is always smaller than $\Delta t_{\rm sound}$, and \citet{Shen_2018a} limited the energy release during $\Delta t_{\rm hydro}$ in a similar way,
the value of $|\Delta \ln T|_{\rm max}$ is larger here for the same energy release rate limit as in \citet{Shen_2018a}.

Having an energy release rate that is independent of timestep choice ameliorates one of the features of our limiter implementation criticized by \cite{Kushnir_Katz_2020}.
Their method also goes further to implement additional limiter criteria, for example with some treatment of composition, that we do not pursue here.
We believe that our current limiter implementation is sufficient for our desired accuracy under the methods utilized here.
Temperature histories are post-processed to obtain the final yields and there is no explicit treatment of detonation curvature, which is sensitive to the thickness of the reaction front.

The energy limit, including calculation of $c_V$, is evaluated at the beginning of the timestep.
The nuclear reactions are then integrated using a sub-stepped Bader-Deuflhard scheme until a time $\Delta t_{\rm hydro}$ is reached or the energy limit is overstepped.
In the latter case, the final abundances for the step are determined by linear interpolation between the last two sub-steps of the integration, such that the energy release is the limit $\Delta \mathcal E$.

\subsection{Domain fluff}

The portion of the computational domain outside the initial star is filled with low-density material.
We refer to this material, as is conventional, as ``fluff.''
While this material is low enough density to be a negligible portion of the overall mass on the domain, some care in choosing its state is necessary in order to not adversely impact the simulation.
Our goal is for the fluff to have as minimal effect on the explosion and ejecta computation as possible and we do not attempt to model any sort of circumstellar medium.
There are two significant issues to be addressed in setting the state, which are in tension with one another.
The first of these implies that the density should be chosen as small as possible, while the second limits how small the density can be.

The lowest possible density is desirable because, for any given fluff density, the outgoing ejecta will eventually encounter enough fluff that a reverse shock will be created at some finite ejecta speed.
Thus a higher density fluff limits the time to which a simulation can be run without the reverse shock falling below some desired speed in the ejecta such that it affects a substantial amount of ejecta mass.
For example, in previous work \citep{Townsley_2019a}, and the cases with masses of 0.85 and $1.02 \, M_\odot$ with shell base density $2 \times 10^5$ g~cm$^{-3}$ here, a fluff density of $10^{-3}$~g~cm$^{-3}$ limited the time of the simulation to about 25 seconds, by which some regions of slower eject in the southern hemisphere had a reverse shock below 20,000~km~s$^{-1}$.
(See Figure 2 in \citealt{Townsley_2019a}.)
This limitation has been present in many of the simulations based on the setups introduced by \citet{Townsley_etal_2009}.

The second issue prevents the density from being chosen arbitrarily small.
The sound speed in the fluff must be kept from being so large that the sound crossing time of cells in the fluff region do not constrain the timestep to be smaller than it otherwise would be based on the star's interior.
Adjacent blocks (regions of the grid that must have a uniform cell size, typically $8\times 8$ or $16\times 16$ cells in size) can only differ by a factor of 2 in cell size.
As a result, the smallest cell size in the fluff is not appreciably different from that in the star and ejecta.
Since the sound speed in a low-density, radiation-dominated fluid scales as $\sqrt{T^4/\rho}$, this creates a lower limit on the fluff density.

The sound speed constraint also creates an indirect lower limit on the temperature of the fluff near the surface of the initial hydrostatic star.
The outer edge of the star is, by necessity, unresolved on the computational mesh.
For any stellar surface, the scale height, which is the length on which the density changes by order unity, becomes very small compared to the overall stellar size near the photosphere.
This defines the ``edge'' of the star, where the density drops off to zero.
Since this region is unresolved in our initial state, the last cell containing stellar material (not just fluff) has a finite, and usually appreciable, pressure.
With no resolution limitation, this pressure would be balanced by gravity.
However, the limited resolution does not allow that.
As a result, if the temperature of the fluff is chosen too low so that the pressure in the fluff (which is almost entirely radiation pressure, due to the low density) is much lower than the pressure of the last zone of the star, a shock will be created in the fluff.
This shock will cause the temperature in the fluff to rise quite significantly due to the low density, leading to a sound speed that is too high.
As such, the minimum temperature of the fluff near the star's surface, and therefore the minimum density as well, is determined by the resolution of the grid at the stellar surface.
A finer spatial grid means the edge of the star has a lower pressure and therefore the fluff can be lower temperature and density and still have a low enough sound speed.

Unfortunately the minimum density implied by the limitation on the sound speed and the finite resolution at the edge of the star is such that the reverse shock will appear at around 20,000~km~s$^{-1}$ in around 20 seconds or so for a typical supernova ejecta.
This problem can be avoided by introducing an outward down-gradient in both the density and temperature together, maintaining $T^4/\rho$ at a uniform value.
By making this gradient physically large in extent, any dynamics it introduces due to the concomitant gradient in pressure is slow enough to not be relevant to the simulation.

We make choices of the fluff profile that, in testing, proved sufficient to keep the reverse shock outside the location marked as fluff using a composition-like mass scalar for simulations of up to 100 seconds.
As a result, the velocity extent of the usable ejecta is now determined by the resolution of the grid at the star-fluff boundary.
For this work we choose to start the decrease in the $T$ and $\rho$ of the fluff at the star-fluff boundary, decreasing in a log-linear fashion to a radius of $5\times 10^{10}$ cm, outside of which the density is $10^{-11}$~g~cm$^{-3}$ and the temperature $3 \times 10^{5}$ K.
As mentioned above, the $T$ and $\rho$ near the edge of the star is determined by that necessary to be similar to the pressure in the last zone of the star.
Specifically, we set the temperature of the fluff at the edge of the star such that the radiation pressure in the fluff is equal to the gas pressure of the last zone of the star.
The inner density is such that the sound speed is approximately constant throughout the fluff at about $2\times 10^9$~cm~s$^{-1}$.

\subsection{Post-processing for nucleosynthesis}
To enable more accurate nuclear yields, a particle tracing post-processing technique is used as in \citet{Shen_2018a} and \citet{Townsley_2019a} following completion of the primary hydrodynamic nuclear simulation.
At initialization, approximately 100,000 Lagrangian particles are distributed throughout the core and shell of the star.
In order to enable sufficient sampling of the shell ashes where the ejecta vary the most, 20\% of the particles are assigned to the He shell.
In addition to having more particles per mass in the shell, a region of the shell within $10^\circ$ of the symmetry axis is chosen for even higher density of particles, with 20\% of the shell particles designated to this area.
The particles are of equal mass within their respective region and are distributed randomly, weighted by mass in an effort to properly sample the entire star.
A single shell particle in our thin-shell models represents at most $2.5 \times 10^{-6}$ M$_\odot$.

The position, velocity, temperature, and density is recorded for every particle at each timestep in the simulation.
\texttt{MESA}'s one zone burner with a 205-nuclide nuclear network is used with the temperature and density history of each particle to determine the corresponding elemental evolution.
\cite{Shen_2018a} used the same post-processing scheme for similar detonation simulations and found that nuclear yields between the main simulation and post-processing stages differ by upwards of 10\% (for a less complete, 41-isotope network), with energetics being consistent within 0.3\%.
\citet{Miles_etal_2019} showed that this method also gives good agreement with results in which the unresolved detonation structure is explicitly reconstructed.
Combining the final position, velocity, and mass fraction of each particle with the final state of the simulation, we generate a 2-dimensional ejecta profile.

\subsection{Simulations}
\label{subsec:simulations}

\begin{figure*}
\includegraphics{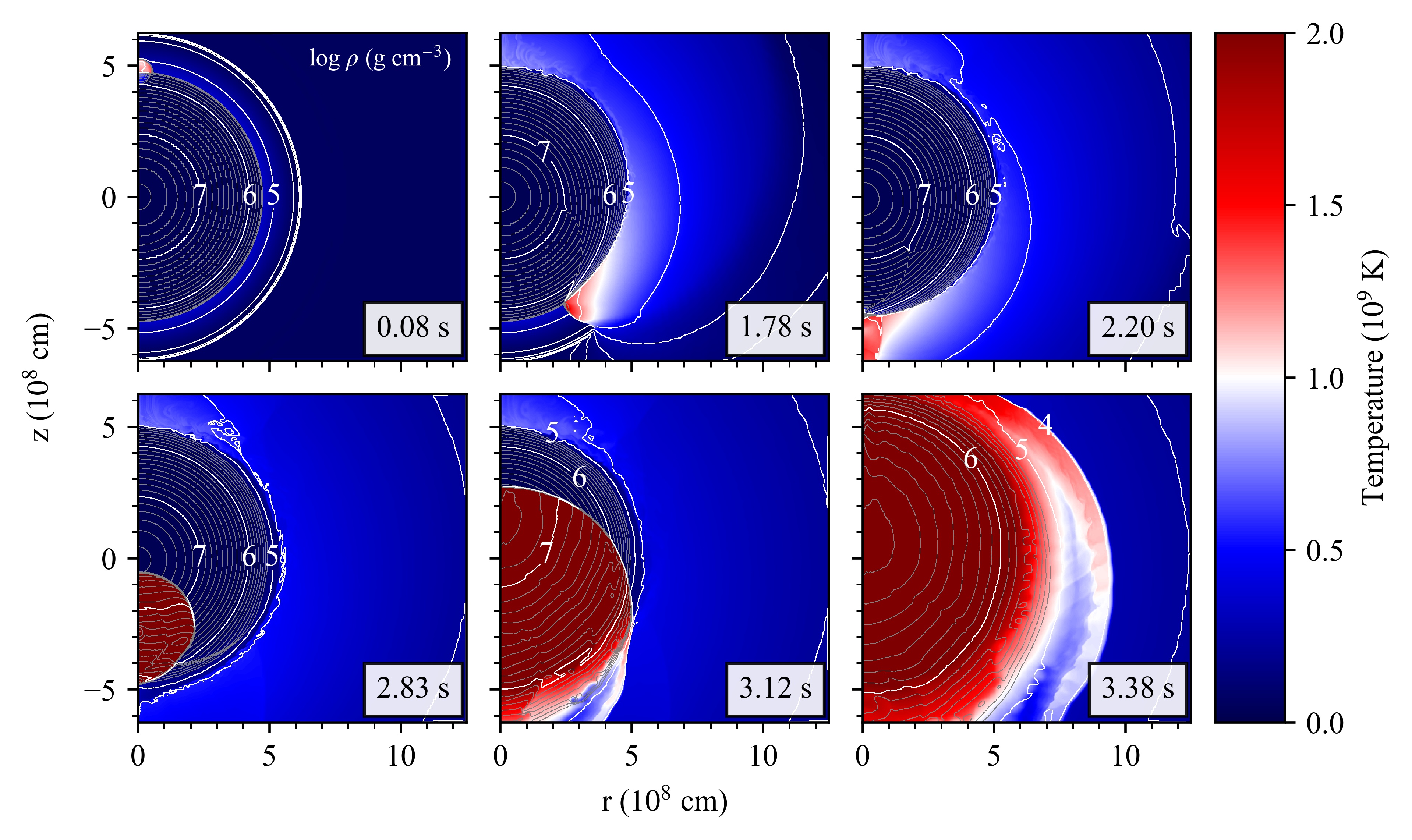}
\centering
\caption{
Temperature and density during a double detonation explosion simulation of a $1.0 \, M_\odot$ C-O WD with a thin $0.016 \, M_\odot$ He shell. 
Temperature is shown as a color scale, while white and grey lines are logarithmically equally spaced density contours.
White contours are spaced every decade and labeled for the three highest values in each frame, while grey contours are only shown for $\log \rho = 5.3$ (the initial density at the core-shell interface) and above and are spaced to give 10 intervals per decade.
}
\label{fig:walkthrough}
\end{figure*}

2-dimensional full-star detonation simulations were conducted using the modular multi-physics code \texttt{FLASH} version 4.3.
The built-in nuclear energy deposition in \texttt{FLASH} is replaced with reaction integration using \texttt{MESA} described above in Section \ref{sec:network}.

Our simulations utilize an adaptive mesh in \texttt{FLASH}.
With this, regions with large temperature, density, or pressure gradients and energy generation are increased in resolution. 
The maximum resolution of the simulations is typically 4 km.
Regions outside the star are not refined.
The strategy utilized is as described in detail in \citet{Townsley_etal_2009}.
The sub-Chandrasekhar-mass WD progenitors used here are about a factor of two larger in radius than the Chandrasekhar-mass progenitors used in \citet{Townsley_etal_2009}.
Thus the resolution of non-burning regions, needed to maintain hydrostatic balance of the progenitor, is 32 instead of 16 km.
The exception to this is that full resolution is enforced at the core-shell composition boundary.
The location of the detonation in the He shell is tracked and this enhanced resolution condition is relaxed after the detonation has passed a given region.
Material that expands beyond the radius at which the initial stellar surface is located has a minimum cell size of 64 km.

Once the burning phases, both He and C detonations, of the explosion are complete, the grid is coarsened to a uniform maximum refinement level that coarsens as the ejecta expands.
The minimum cell size is chosen so that there are no more than 1024 cells between the center of the domain and the edge of the expanding shell.
This yields several changes in maximum refinement level between the end of the burning at about 3 seconds after He ignition and the end of the simulation at 100 seconds.

To initiate the shell detonation phase, a circular isochoric hotspot is placed on the core-shell interface.
The temperature profile declines linearly from the center of the hotspot, with a maximum temperature of $1.0 \times 10^9$ K and the minimum temperature matching that of the base layer of the ambient He shell.
The hotspot radii used range between 100 - 500 km and vary across the progenitors as they were chosen to be the smallest necessary, within 50 km, to trigger a He shell detonation.
Thinner He shells require larger hotspots to trigger a detonation.
We note that while our ignition scheme here represents the single ignition point scenario, the multi-point ignition scenario has been shown in 3D to also robustly trigger a core detonation \citep{moll_multi-dimensional_2013}.

In order to properly capture the convergence of the inward traveling shock from the He shell detonation, refinement of the shock region within the WD is forced until ignition is reached.
In a single run of the progenitor with the least massive shell, the maximum refinement was temporarily improved to 2 km in order to observe a core detonation due to the convergence of the inward shock.
The limiter was relaxed briefly during both ignition phases as the main goal of this study is to characterize the overall explosion outcome, leaving detailed study of the ignition processes to separate work.

Most of our simulated explosions were allowed to run for 100 seconds from primary ignition, the large majority of which the ejecta spends in a generally homologous state.
Three of our runs utilized an earlier fluff scheme that only allowed the simulation to run to around 25 seconds.
There are no consequential differences between ejecta profiles in velocity space due to these differing running times, only in the velocity extent of the usable ejecta region due to the location of the reverse shock.

The 2-dimensional ejecta profiles are constructed on a uniform grid in velocity.
We use a velocity cell size of $500\times 500$ km~s$^{-1}$.
The scaling from the spatial grid in the simulation to the velocity grid is determined by averaging the relation $t_{\rm exp} = r/v_r$ over all the final states for the tracer particles.
Here $v_r$ is the radial component of the velocity of the material (as sampled by the particle) and $r$ is the radial location of each particle.
$t_{\rm exp}$ is the resulting expansion time, which is the time since the homologous (i.e. Hubble-flow-like) expansion started assuming a time-independent velocity.
These are shown for each simulation in Table \ref{table:parameters} and are generally a few seconds shorter than the simulation time due to the time required for the energy deposition prior to ejection.
Once $t_{\rm exp}$ is determined, it can be used to map any spatial location to an equivalent velocity location.

The spatial density distribution from the hydrodynamic simulation is then mapped to the velocity grid and regridded (averaged or interpolated) onto the uniform velocity grid.
The abundances are then determined by averaging the results of the individual tracer particle nucleosynthesis using a cloud-in-cell method based on the final velocity for each particle.
Cells in which the mass scalar marking the fluff (non-star) material in the hydrodynamic simulation is greater than 1\% or for which $v_r < r/t_{\rm exp}-500$~km~s$^{-1}$, the latter being due to
the impact of the reverse shock, are excluded from the ejecta and filled with low density pure He.
In the outer regions of the ejecta, some cells may not contain any tracer particles.
For these, the closest cells that do contain particles are determined and abundances are interpolated where multiple cells are available or extrapolated using a constant outward profile.

Many species present at 100 s will decay by the time of maximum light.
Except as indicated when a specific isotope is designated, we here show total elemental abundances after 12 days of decay using rates included in the tables available in \texttt{MESA}.
Both single- and two-step decay chains are computed analytically, which is sufficient for the 205-nuclide set used for nucleosynthetic post-processing.
The following elements are not decayed: $^{56}$Ni, $^{56}$Co, $^{52}$Fe, $^{48}$Cr, and $^{48}$V.
These are left to be treated explicitly in any follow-up radiative transfer computations.
$^{52}$Mn, produced to the ground state in our simulations, is left to decay.

\section{Results}
\label{sec:results}

We have simulated double detonations in eleven unique progenitors and determined the yields and ejecta profiles for each.
This section contains a presentation of our results as well as a comparison to previous work and discussion of particular highlights.

\subsection{Overview of simulation}
We first give a brief overview of the explosion process that occurs in a similar way in each simulation.
Figure \ref{fig:walkthrough} shows representative times from the double detonation explosion simulation of a $1.0 \, M_\odot$ C-O WD with a thin $0.016 \, M_\odot$ He shell. 
This figure also demonstrates the assumed azimuthal symmetry, such that the domain is a 2-dimensional plane whose left edge ($r=0$ here) is the symmetry axis.
Azimuthal symmetry is appropriate under the assumption we have made here of a single ignition site.
We will refer to the $+z$ hemisphere, where the He ignition occurs, as the ``northern'' hemisphere, and the $-z$ as the ``southern''.

\begin{figure}
\centering
\includegraphics{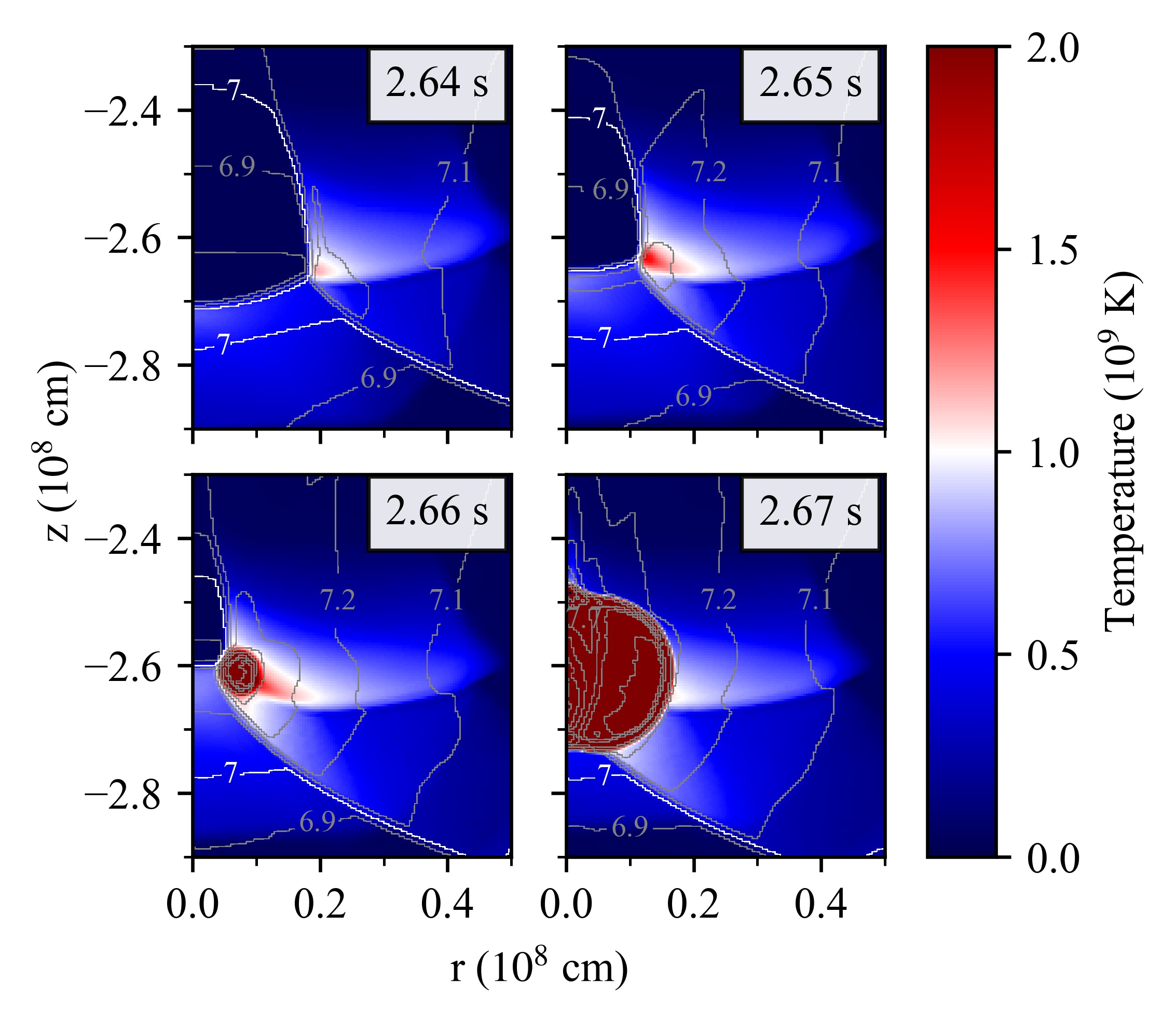}
\caption{Zoomed-in frames showing the temperature and density during the secondary, core ignition in our simulation of a $0.85 \, M_\odot$ WD with a thin He shell of $0.033 \, M_\odot$.
Temperature and density are shown as a color scale and contours, respectively, as in Figure \ref{fig:walkthrough}.
The point of first ignition is located well away from the symmetry axis in the simulation, the left edge of the domain.
}
\label{fig:secondary_ignition}
\end{figure}

\begin{figure}
\centering
\includegraphics{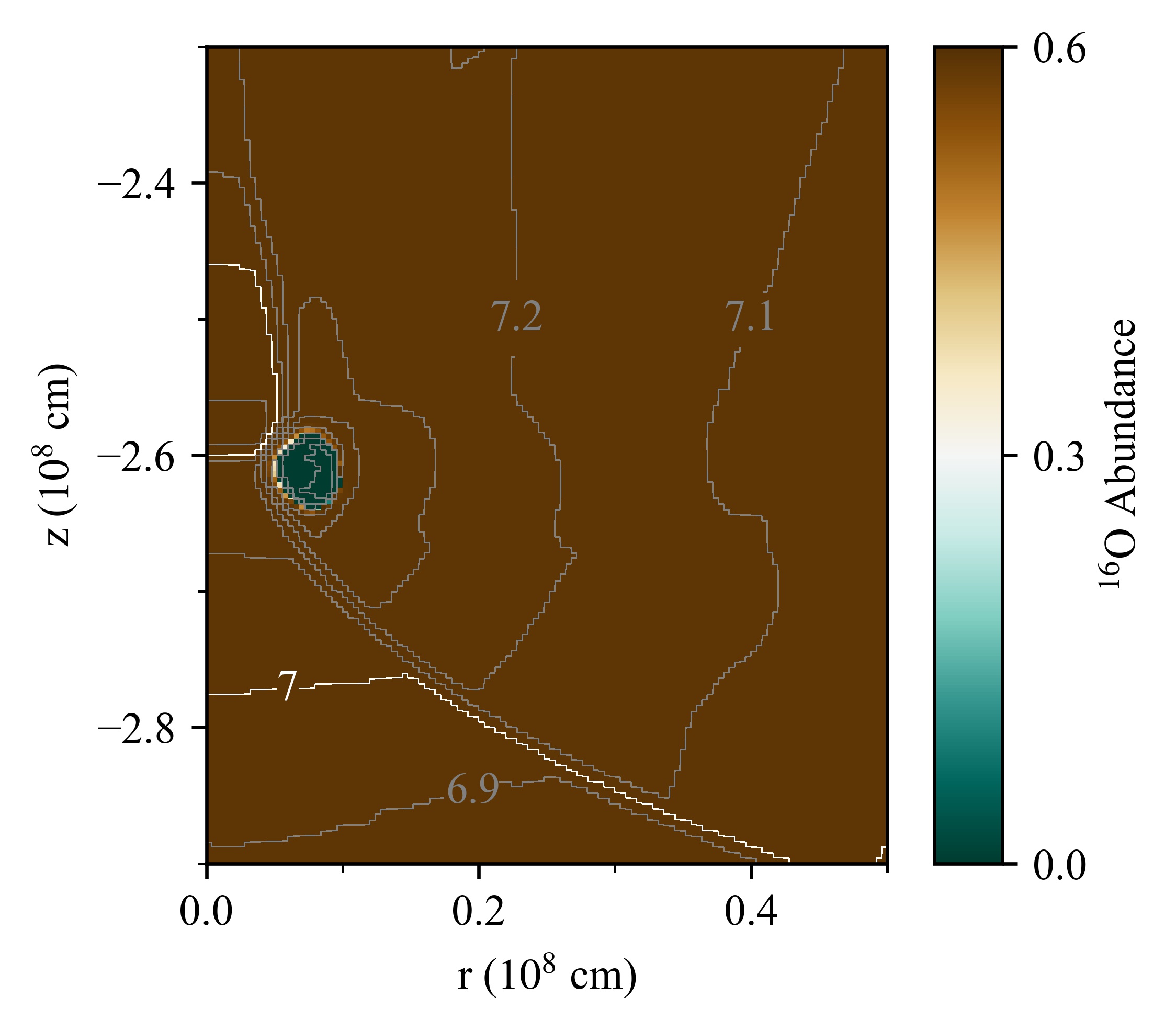}
\caption{Same as the 2.66 s frame in Figure \ref{fig:secondary_ignition} but for $^{16}$O abundance (mass fraction) instead of temperature, showing the core ignition is clearly off the symmetry axis.
}
\label{fig:secondary_ignition_o}
\end{figure}

In the first frame, the He burning can be observed growing from a small hotspot placed at the core-shell interface.
At 1.78 s, the He shell detonation is nearing completion and the inward-traveling shock in the southern hemisphere can be observed in the distortion of the core density contours.
This shock can be seen closer to the central axis at 2.20 s after the shell He has been depleted.
The convergence of this shock and associated secondary ignition are detailed in Figures \ref{fig:secondary_ignition} and \ref{fig:secondary_ignition_o}.
By 2.83 s the core, secondary detonation has been triggered and C-O detonation proceeds.
In the last frame, all fuel has been burned and the star is expanding due to the deposited energy.
After this point, the simulation is allowed to evolve to 100 seconds, well into a time where effective ejecta homology is reached.

\subsection{Dynamics of Secondary Ignition}

Since our main goal is to obtain ejecta that can be compared to observed objects, we did not perform a detailed investigation of the secondary ignition.
However, it is worthwhile to comment on what we observe in our simulation and how it relates to limitations due to resolution and the imposed axisymmetry. 
We find, in at least some of our models, a core detonation ignition that takes place some distance from the symmetry axis (see Figures \ref{fig:secondary_ignition} and \ref{fig:secondary_ignition_o} for an example).
This is in one of our thin-shell models, with the resulting C detonation ignition point around 7.5 $\times\ 10^{6}$ cm from the symmetry axis.
This feature mimics the ignition dynamics seen in analytical work on 1-dimensional converging shock ignitions \citep{Kushnir_etal_2012,Shen_2014a}, but bears further study in multiple dimensions particularly to address a more realistic, possibly multi-point, ignition geometry.

In all cases, the trailing edge of the converging shock grew to temperatures of over $10^{9}$ K, but did not always lead to an obvious ignition before convergence along the symmetry axis.
While most prominent in the higher-mass shell cases, we find that a clearly off-axis core ignition does not necessarily occur in all of our runs. 
It is difficult to determine precisely if the ignition occurs off-axis in the unclear cases due to spatial resolution issues.
Additionally, increasing the resolution in these cases would only increase the effect that the symmetry axis has on producing an ignition.

The only progenitor that had a systematically unique detonation was our thickest-shell case ($M_{\rm tot} = 1.0 \, M_\odot$, $M_{\rm sh} = 0.1 \, M_\odot$).
In it, the second detonation is triggered where the converging shock meets the core-impacting shell material as the He detonation is completing and colliding near the south pole, occurring slightly earlier and closer to the core-shell boundary than in our other nine runs.
This may to be similar to the “scissor” mechanism (coined by \citealt{gronow_sne_2020}) seen in other thick-shell, multi-dimensional simulations \citep{garcia_senz_2018,gronow_sne_2020}.
In \citet{gronow_sne_2020}, this alternative core ignition ultimately had little significant effect on the calculated observables between similar models.
While the precise location and mechanism behind this secondary ignition is different to our other runs, we still present the yields and ejecta from this case as useful comparisons.
The aspects of this core ignition in the thick shell case are particularly sensitive to the use of a limiter and choice of resolution and thus leave detailed analysis to separate work.

\subsection{Yields}

\begin{table*}
\centering
\caption{Explosion Yields 1}
\begin{tabular}{ccc|cccccccc}
M$_{tot}$ & $\rho_{b,5}$\footnote{Shell base density ($10^{5}$ g cm$^{-3})$} & M$_{shell}$ & LME & Shell LME & IME & Shell IME & HME (-$^{56}$Ni) & Shell HME (-$^{56}$Ni) & $^{56}$Ni & Shell $^{56}$Ni \\ 
($M_\odot$) &  & ($M_\odot$) & ($M_\odot$) & ($M_\odot$) & ($M_\odot$) & ($M_\odot$) & ($M_\odot$) & ($M_\odot$) & ($M_\odot$) & ($M_\odot$) \\ \hline \hline
0.85 & 2 & 0.033 & 0.213 & 0.023 & 0.499 & 9.7$\times 10^{-3}$ & 0.027 & 8.1$\times 10^{-5}$ & 0.108 & 5.6$\times 10^{-6}$ \\ 
0.90 & 2 & 0.025 & 0.172 & 0.018 & 0.457 & 7.1$\times 10^{-3}$ & 0.036 & 5.4$\times 10^{-5}$ & 0.236 & 4.2$\times 10^{-6}$ \\ 
1.00 & 2 & 0.016 & 0.102 & 0.012 & 0.348 & 4.4$\times 10^{-3}$ & 0.055 & 4.0$\times 10^{-5}$ & 0.495 & 1.9$\times 10^{-6}$ \\ 
1.10 & 2 & 8.4$\times 10^{-3}$ & 0.059 & 6.4$\times 10^{-3}$ & 0.220 & 2.0$\times 10^{-3}$ & 0.075 & 6.4$\times 10^{-5}$ & 0.747 & 8.7$\times 10^{-7}$ \\ \hline
1.02 & 2 & 0.021 & 0.094 & 0.015 & 0.339 & 6.1$\times 10^{-3}$ & 0.058 & 4.1$\times 10^{-5}$ & 0.530 & 6.8$\times 10^{-7}$ \\ 
1.02 & 2 & 0.021\footnote{Decreased $^{16}$O} & 0.103 & 0.017 & 0.327 & 4.3$\times 10^{-3}$ & 0.058 & 4.7$\times 10^{-5}$ & 0.534 & 4.5$\times 10^{-7}$ \\ \hline
0.85 & 3 & 0.049 & 0.191 & 0.030 & 0.486 & 0.017 & 0.030 & 1.4$\times 10^{-3}$ & 0.144 & 3.2$\times 10^{-5}$ \\ 
1.00 & 3 & 0.021 & 0.098 & 0.014 & 0.340 & 7.5$\times 10^{-3}$ & 0.057 & 3.0$\times 10^{-4}$ & 0.507 & 8.2$\times 10^{-6}$ \\ 
1.10 & 3 & 0.011 & 0.058 & 7.4$\times 10^{-3}$ & 0.212 & 3.9$\times 10^{-3}$ & 0.075 & 1.1$\times 10^{-4}$ & 0.755 & 3.3$\times 10^{-6}$ \\ \hline
1.00 & 6 & 0.042 & 0.098 & 0.020 & 0.310 & 6.5$\times 10^{-3}$ & 0.071 & 0.015 & 0.522 & 8.3$\times 10^{-4}$ \\ 
1.00 & 14 & 0.100 & 0.107 & 0.030 & 0.257 & 3.0$\times 10^{-3}$ & 0.074 & 0.019 & 0.562 & 0.047 \\ 
\end{tabular}
\label{table:abund1}
\end{table*}

\begin{sidewaystable}
\centering
\caption{Explosion Yields 2}
\begin{tabular}{ccc|cccccccccc}
M$_{tot}$ & $\rho_{b,5}$\footnote{Shell base density ($10^{5}$ g cm$^{-3})$} & M$_{shell}$ & $^{12}$C & Shell $^{12}$C & $^{28}$Si & Shell $^{28}$Si & $^{40}$Ca & Shell $^{40}$Ca & $^{44}$Ti & Shell $^{44}$Ti & $^{48}$Cr & Shell $^{48}$Cr \\ 
($M_\odot$) & ($10^{5}$ g cm$^{-3})$ & ($M_\odot$) & ($M_\odot$) & ($M_\odot$) & ($M_\odot$) & ($M_\odot$) & ($M_\odot$) & ($M_\odot$) & ($M_\odot$) & ($M_\odot$) & ($M_\odot$) & ($M_\odot$) \\ \hline \hline
0.85 & 2 & 0.033 & 0.014 & 1.9$\times 10^{-4}$ & 0.282 & 1.1$\times 10^{-3}$ & 0.023 & 1.2$\times 10^{-3}$ & 3.3$\times 10^{-5}$ & 2.1$\times 10^{-5}$ & 2.7$\times 10^{-4}$ & 6.3$\times 10^{-6}$ \\ 
0.90 & 2 & 0.025 & 0.011 & 1.8$\times 10^{-4}$ & 0.255 & 8.4$\times 10^{-4}$ & 0.023 & 6.1$\times 10^{-4}$ & 2.5$\times 10^{-5}$ & 1.1$\times 10^{-5}$ & 3.8$\times 10^{-4}$ & 3.7$\times 10^{-6}$ \\ 
1.00 & 2 & 0.016 & 4.1$\times 10^{-3}$ & 1.3$\times 10^{-4}$ & 0.191 & 5.5$\times 10^{-4}$ & 0.020 & 3.5$\times 10^{-4}$ & 3.2$\times 10^{-5}$ & 1.1$\times 10^{-5}$ & 4.0$\times 10^{-4}$ & 2.7$\times 10^{-6}$ \\ 
1.10 & 2 & 8.4$\times 10^{-3}$ & 1.9$\times 10^{-3}$ & 9.7$\times 10^{-5}$ & 0.119 & 3.1$\times 10^{-4}$ & 0.014 & 2.4$\times 10^{-4}$ & 5.7$\times 10^{-5}$ & 3.4$\times 10^{-5}$ & 3.4$\times 10^{-4}$ & 1.2$\times 10^{-5}$ \\ \hline
1.02 & 2 & 0.021 & 2.9$\times 10^{-3}$ & 1.4$\times 10^{-4}$ & 0.184 & 7.0$\times 10^{-4}$ & 0.020 & 5.0$\times 10^{-4}$ & 2.9$\times 10^{-5}$ & 8.3$\times 10^{-6}$ & 4.1$\times 10^{-4}$ & 1.4$\times 10^{-6}$ \\ 
1.02 & 2 & 0.021\footnote{Decreased $^{16}$O} & 4.2$\times 10^{-3}$ & 2.7$\times 10^{-4}$ & 0.180 & 9.1$\times 10^{-4}$ & 0.018 & 2.3$\times 10^{-4}$ & 3.6$\times 10^{-5}$ & 1.5$\times 10^{-5}$ & 3.8$\times 10^{-4}$ & 1.5$\times 10^{-6}$ \\ \hline
0.85 & 3 & 0.049 & 0.011 & 1.3$\times 10^{-4}$ & 0.268 & 7.5$\times 10^{-4}$ & 0.033 & 0.011 & 1.2$\times 10^{-3}$ & 1.2$\times 10^{-3}$ & 3.7$\times 10^{-4}$ & 7.3$\times 10^{-5}$ \\ 
1.00 & 3 & 0.021 & 3.2$\times 10^{-3}$ & 5.9$\times 10^{-5}$ & 0.185 & 3.6$\times 10^{-4}$ & 0.023 & 4.4$\times 10^{-3}$ & 2.7$\times 10^{-4}$ & 2.5$\times 10^{-4}$ & 3.9$\times 10^{-4}$ & 1.1$\times 10^{-5}$ \\ 
1.10 & 3 & 0.011 & 1.4$\times 10^{-3}$ & 3.2$\times 10^{-5}$ & 0.114 & 2.1$\times 10^{-4}$ & 0.015 & 1.9$\times 10^{-3}$ & 1.1$\times 10^{-4}$ & 8.2$\times 10^{-5}$ & 3.2$\times 10^{-4}$ & 7.8$\times 10^{-6}$ \\ \hline
1.00 & 6 & 0.042 & 2.4$\times 10^{-3}$ & 1.1$\times 10^{-5}$ & 0.168 & 1.0$\times 10^{-4}$ & 0.022 & 5.1$\times 10^{-3}$ & 2.8$\times 10^{-3}$ & 2.8$\times 10^{-3}$ & 6.4$\times 10^{-3}$ & 6.0$\times 10^{-3}$ \\ 
1.00 & 14 & 0.100 & 5.9$\times 10^{-4}$ & 1.2$\times 10^{-6}$ & 0.141 & 1.3$\times 10^{-5}$ & 0.016 & 2.4$\times 10^{-3}$ & 1.3$\times 10^{-3}$ & 1.3$\times 10^{-3}$ & 3.0$\times 10^{-3}$ & 2.7$\times 10^{-3}$ \\ 
\end{tabular}
\label{table:abund2}
\end{sidewaystable}

The main final yields for each simulation are shown in Table \ref{table:abund1}.
Displayed are the post-processed yields at 100 s with each isotope grouped into low- (Z $\leq$ 10, LME), intermediate- (11 $\leq$ Z $\leq$ 20, IME), and high-mass element (Z $\geq$ 21, HME) groups.
The $^{56}$Ni yields are subtracted from the HME and displayed on their own for clearer analysis.
Yields for additional significant isotopes, also post-processed and at 100s, are shown in Table \ref{table:abund2}.
Yields that are attributed to the original shell material are shown in addition to the total yields for each element group and isotope.

Core yields are directly correlated with core mass with more massive progenitors generating more HME, including $^{56}$Ni.
This is a reflection of the more complete burning expected in more massive WD detonations.
Likewise, the IME and LME yields generally decrease with core mass.
This can be observed in greater detail in Table \ref{table:abund2} where the yields of the relatively lower-mass elements ($^{28}$Si, $^{40}$Ca, and $^{44}$Ti) decrease with core mass.
The range of $^{56}$Ni produced in all of our progenitors are consistent with that of normal, observed SNe Ia \citep{stritzinger_2006}.

There appears to be a relationship between shell mass and total HME yields, even though very little HME is generated in the shell itself, in that more massive shells result in slightly more HME being produced in the core.
This can be seen most clearly for the $1.00 \, M_\odot$ total mass progenitors with shell base densities of 2, 3, and $6 \times 10^{5}~{\rm g}~{\rm cm}^{-3} $.
These three progenitors have HME (including $^{56}$Ni) core yields that increase with shell base density even though the core masses actually decrease slightly (up to $0.025 \, M_\odot$) as shell base density increases.
This effect does not appear to be extremely consequential as the HME yields vary at most by about $0.04 \, M_\odot$.

Yields from the shell material are determined mostly by the combination of the shell density, giving the strength of the detonation, and the mass within the lowest scale height of the shell.
The geometric thickness of the shell determines the curvature which also can affect the yields in a way that is less dominant than the density \citep{Moore_etal_2013}.
Thus, more complete burning occurs and more HMEs are produced in the shell for progenitors with higher shell base densities.
Additionally, progenitors with more massive cores (and thus higher curvature at the shell) will see weaker burning.
This is reflected in our yields.
For example, HME and $^{56}$Ni yields, for any choice of total mass (0.85, 1.00, or $1.10 \, M_\odot$), are higher for denser shells.
This trend is true when weighting for shell mass as well.
With regards to the effects of curvature, this is most easily seen when examining the yields for a choice of shell base density across core masses.
For example, the abundance of $^{56}$Ni is higher for the $0.85 \, M_\odot$ progenitor than that of 1.00 and $1.10 \, M_\odot$ for both the 2 and 3 $\times 10^{5}~{\rm g}~{\rm cm}^{-3}$ shell.
Both of these trends are consistent throughout the yields except for a few special cases, including the $^{48}$Cr produced in the 2 $\times 10^{5}~{\rm g}~{\rm cm}^{-3}$, $1.10 \, M_\odot$ case. 
This discrepancy is explained in Section \ref{sec:reprocessing}.

\subsection{Abundance and Ejecta Profiles}
\label{sec:profiles}

\begin{figure*}
\includegraphics{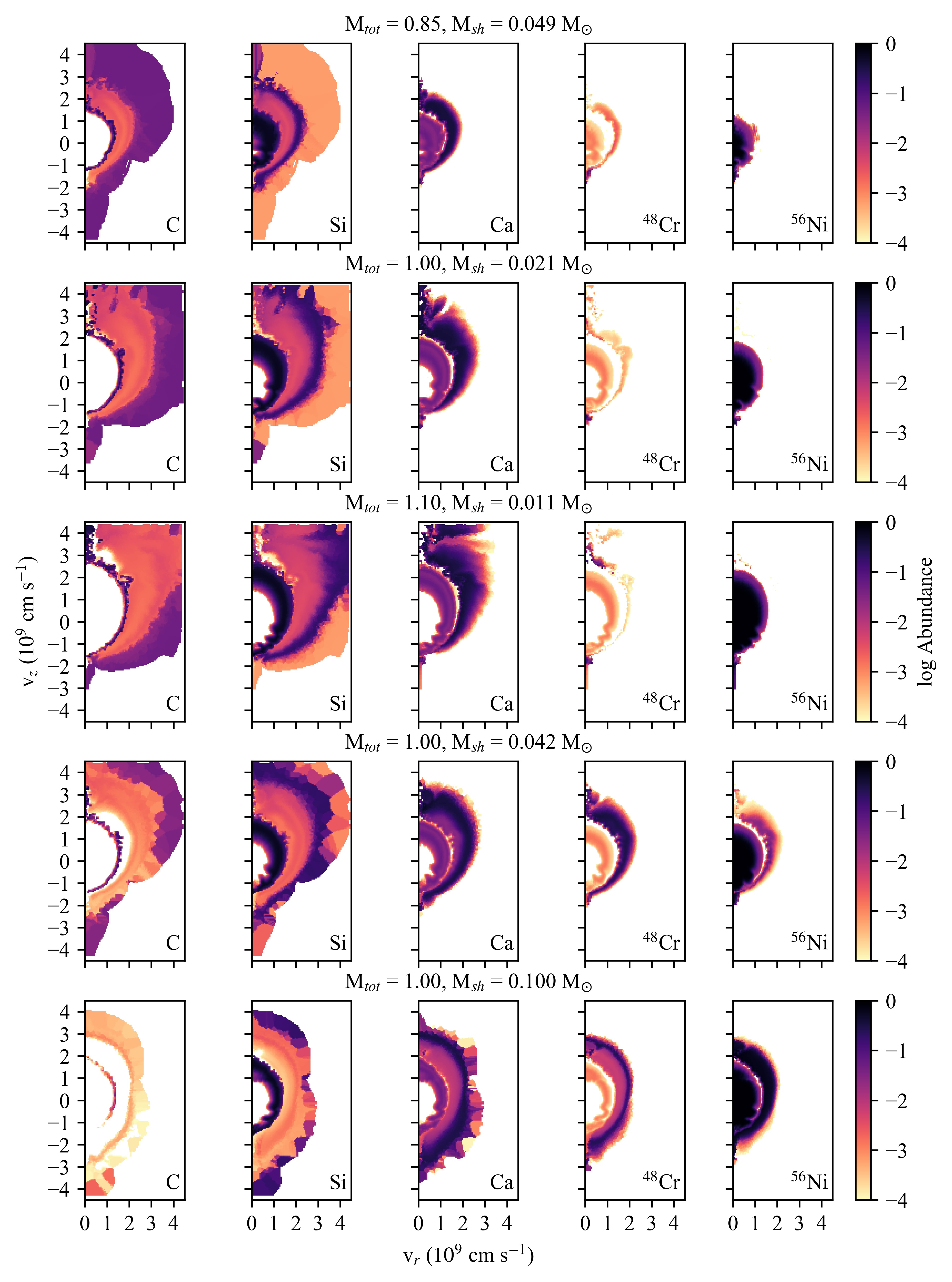}
\centering
\caption{Post-processed and decayed ejecta abundances (mass fraction) for a selection of our models for some consequential elemental/isotopic.
The top three models represent the thin-shell regime while the bottom two are thick-shell models.
A significant amount of radioactive shell material is made in the thick shell models only.
}
\label{fig:2dejecta}
\end{figure*}

\begin{figure*}
\includegraphics{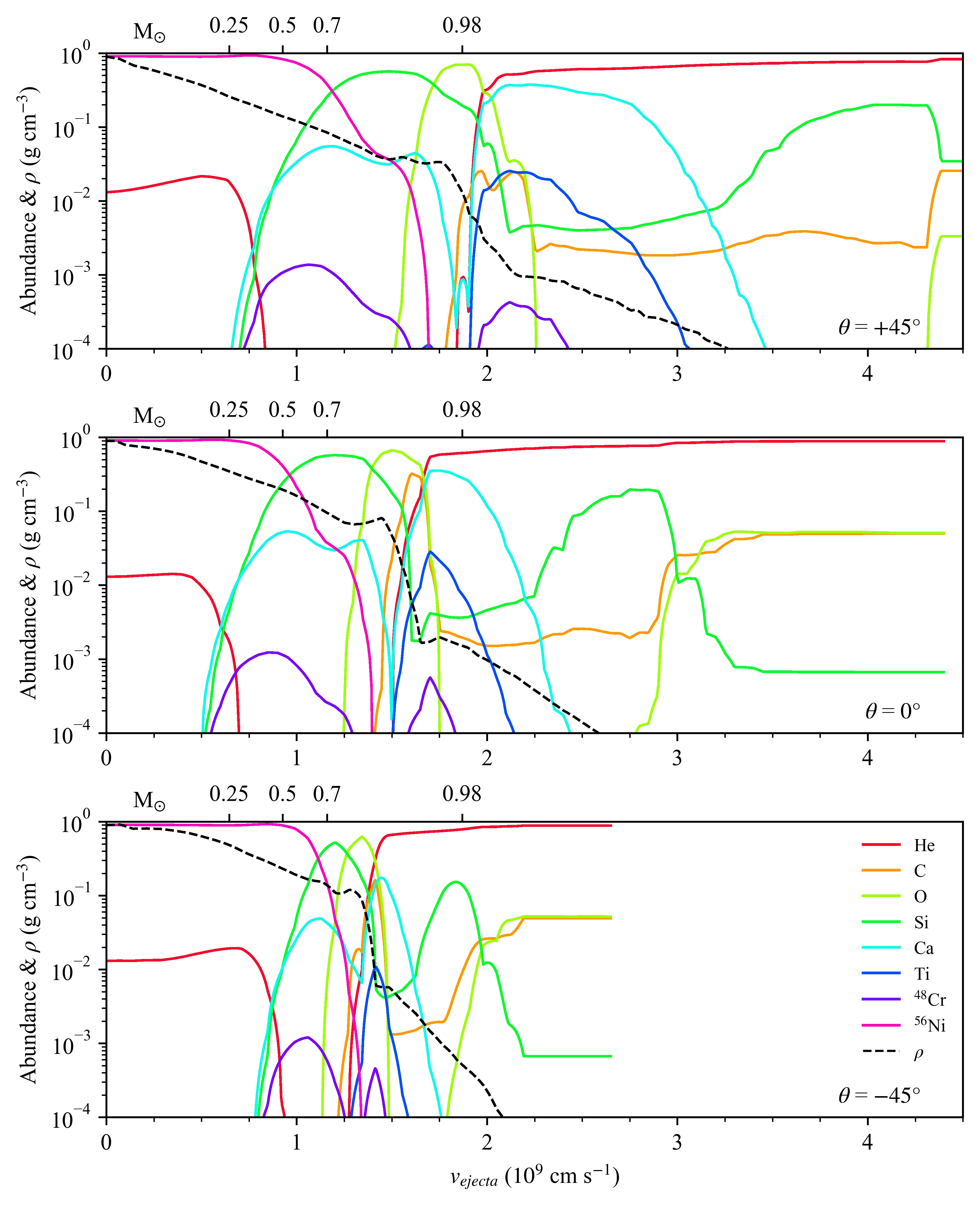}
\centering
\caption{1-dimensional decayed ejecta abundance (mass fraction) profiles for the M$_{tot}$ = 1.00, M$_{\rm sh}$ = 0.021 M$_\odot$ thin-shell model at three angles, $+45^\circ$, $0^\circ$, and $-45^\circ$, relative to the equator.
}
\label{fig:1dprofiles_los}
\end{figure*}

\begin{figure*}
\includegraphics{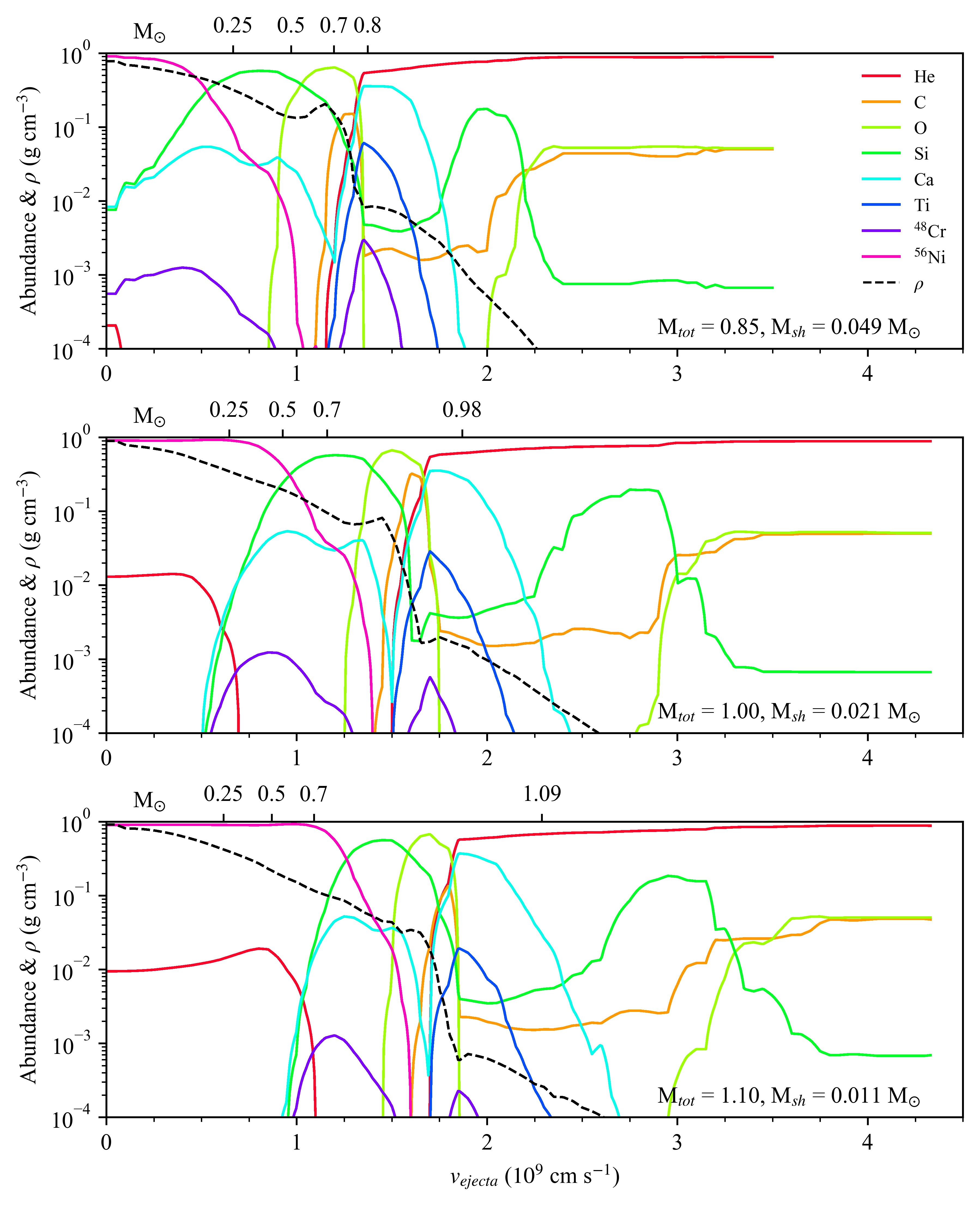}
\centering
\caption{1-dimensional decayed ejecta abundance profiles at the equator for thin-shell progenitors of equal shell base density ($3 \times 10^{5}$ g cm$^{-3}$) but varying core mass.}
\label{fig:1dprofiles_samedens}
\end{figure*}

\begin{figure*}
\includegraphics{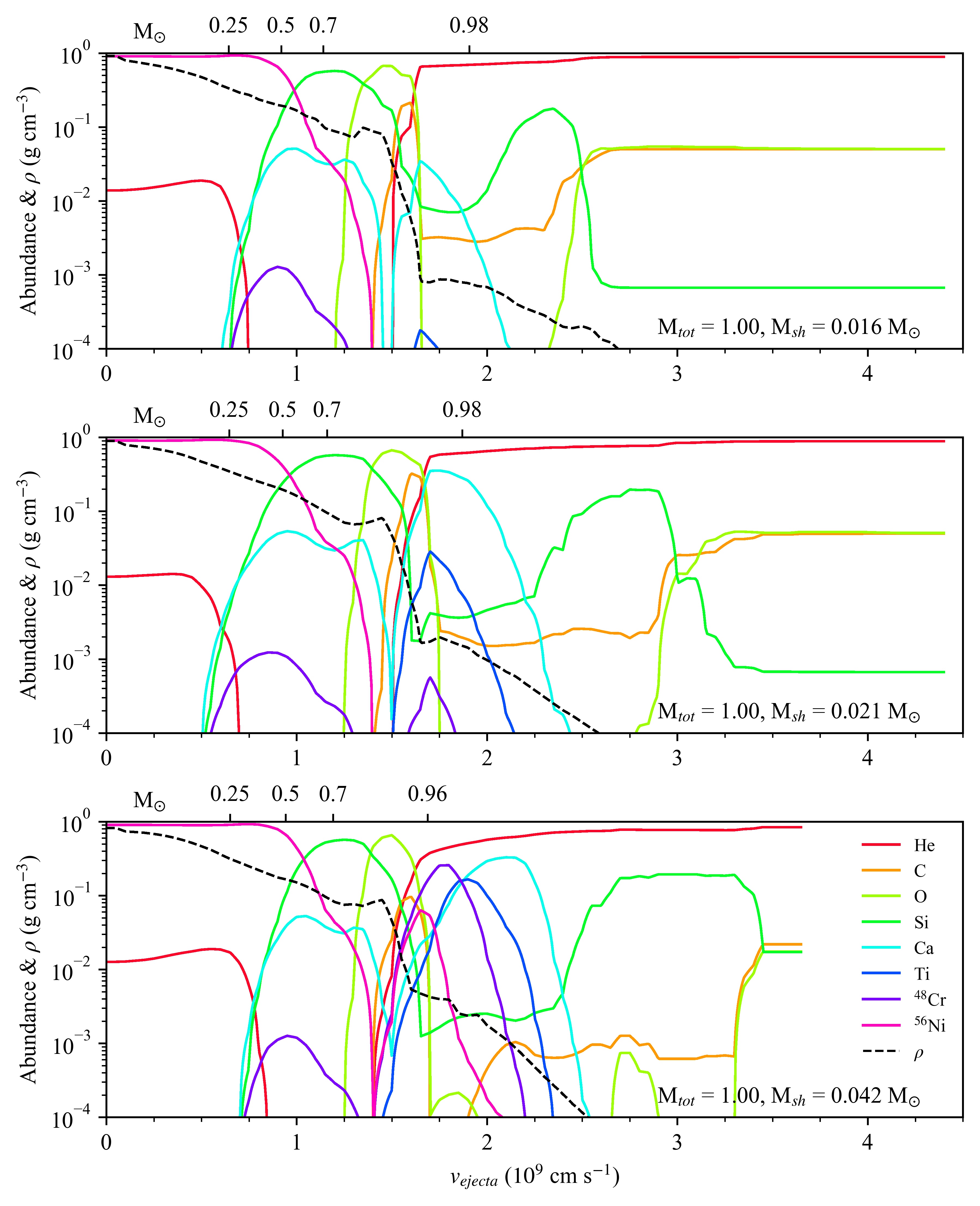}
\centering
\caption{1-dimensional decayed ejecta abundance profiles at the equator for runs of equal total progenitor mass but varying shell thickness.
The thick-shell model presented in the bottom frame produces a significant amount of radioactive material at the base of the shell.
}
\label{fig:1dprofiles_samemass}
\end{figure*}

2-dimensional ejecta from post-processed data for representative progenitors can be seen in Figure \ref{fig:2dejecta}.
1-dimensional ejecta profiles can be seen for a variety of angles in Figure \ref{fig:1dprofiles_los} and for a range of progenitors in Figures \ref{fig:1dprofiles_samedens} and \ref{fig:1dprofiles_samemass}.
The ejecta presented in these figures have been decayed to 14 days, the typical maximum light for observed SNe Ia, as described in Section \ref{subsec:simulations}.

As observed in Figures \ref{fig:2dejecta} and \ref{fig:1dprofiles_los}, the abundance profiles, especially in the outer regions, are very dependent on the polar angle.
Specifically, material generated closer in polar angle to the ignition pole is later found at higher velocities in the ejecta.
For example, Figure \ref{fig:1dprofiles_los} shows the Ti peak at approximately 2.1, 1.7, and 1.4 $\times 10^9$ cm s$^{-1}$ for 1-dimensional ejecta profiles at $+45^\circ$, $0^\circ$, and $-45^\circ$ from the equatorial plane, respectively.
The abundance of the burned products and their velocity widths also varies with polar angle.
This can also be observed in the Ti peaks in Figure \ref{fig:1dprofiles_los} where Ti peak abundance and band velocity width increases with polar angle.
Additionally, the density profiles are asymmetric, with sharper profiles observed in the southern plane.
This results in denser regions of ejected material in the southern plane compared to the corresponding, same initial density material in the northern plane.

The origin of the asymmetric ejecta comes from the systematically off-centered secondary ignitions in our simulations.
The secondary ignition takes place approximately halfway between the core and surface which results in a stronger detonation towards the original ignition point.
Thus, material generated along any given radius in the unperturbed progenitor will have a polar angle dependence in its corresponding ejecta distribution.
The asymmetric nature of our ejecta has been seen before in previous multi-dimensional studies of double detonations \citep{Fink_10,garcia_senz_2018,gronow_sne_2020} and is characteristic of similarly ignited double detonations.

In addition to the bulk yields, the velocity extents of the burned products in the core are clearly influenced by the choice of progenitor with larger cores resulting in faster outer core and shell ashes.
This can be observed in the 1-dimensional abundance profiles at the equator for three separate runs in Figure \ref{fig:1dprofiles_samedens}.
Additionally, we find no obvious visible effect from shell thickness on the inner region of the ejecta.
For example, the distribution of burned core material is mostly unchanged for the progenitors in Figure \ref{fig:1dprofiles_samemass} where the core masses are consistent within $0.025 \, M_\odot$.

The final tick on the mass coordinate axis in Figures \ref{fig:1dprofiles_los}, \ref{fig:1dprofiles_samedens}, and \ref{fig:1dprofiles_samemass} indicates where the enclosed mass at the corresponding velocity is equal to the original core mass.
Notably, this tick is not directly consistent with the position of the products originating from the densest part of the He layer as might be typical in a 1-dimensional or otherwise symmetric simulation (see Figure 3 in \cite{woosley_2011} or Figure 1 in \cite{Polin_2019}).
This is for two reasons: the aforementioned asymmetry of the ejecta and some degree of mixing between core- and shell-burned material.
Inspection of the evolution of abundance profiles and the core-shell interface shows that this mixing occurs mostly during the late burning and early ejecta stage.
Kelvin-Helmholtz (shear) instabilities are generated along the surface of the core as the initial detonation proceeds along the shell.
The velocity imparted to the shell by the detonation causes shell material to flow along the surface of the core, so that some core material mixes with the bottom of the shell.
Additional mixing is also introduced near the end of the secondary detonation as it travels back through the shell at an oblique angle.

The location of the core-shell ash boundary is correlated with break in the rough power-law seen in the density profile in Figures \ref{fig:1dprofiles_los}, \ref{fig:1dprofiles_samedens}, and \ref{fig:1dprofiles_samemass}.
For example, in Figure \ref{fig:1dprofiles_samedens}, the densest shell ashes are aligned with the right edge of the density power-law break at roughly 1.3, 1.8, and 1.9 $\times 10^9$ cm s$^{-1}$ in the $0.85$, $1.00$, and $1.10 \, M_\odot$ cases, respectively.
This region of the ejecta represents the slowest region of shell material and sees the most complete burning of the shell ashes due to the high initial density.
The break in the ejecta density profile is due to the unequal region densities of the progenitor and energy releases of the two detonations.
The aforementioned mixing of core and shell ashes leads to a smoothed break over a width on the order of $10^8$ cm s$^{-1}$.
As observed in Figure \ref{fig:1dprofiles_los}, the density profile of the outer material is sharper in the southern plane, in line with the previously described asymmetric features of the ejecta.
The mixed break in the overall density profile is also much thinner in velocity in the southern plane, perhaps in part attributable to the shorter delay time between when each detonation passes through this region.

As seen in the yields, the properties of the shell have a significant influence on the shell ejecta profiles with thicker shells producing more higher-massed elements.
This is most clear in examination of the core-shell interfaces in Figure \ref{fig:1dprofiles_samemass} which presents ejecta profiles of progenitors with similar core masses but varying shell masses.
For the thinnest progenitor in Figure \ref{fig:1dprofiles_samemass}, the Ca levels are low at the interface in addition to very little HME.
In the thicker shells, more Ca is produced in addition to the presence of Ti and radioactive isotopes.
Thicker shells also result in faster outermost shell-burned ejecta.
This can be observed in Figure \ref{fig:1dprofiles_samedens} where the outermost burned ejecta velocity, roughly associated with the Si peak, increases with shell thickness.

Two strong Si peaks are apparent in the ejecta of each of our runs.
These originate from the outer parts of both the core and shell where burning is less complete (e.g. little to no $^{56}$Ni in these regions).
The exact properties of these peaks vary across the progenitors but are generally consistent.
The Si abundance at these peaks reach upwards of 0.5 and 0.1 for the inner and outer peaks respectively.
The velocity where these peaks occur is also highly dependent on the polar angle.
This is best observed across the frames in Figure \ref{fig:1dprofiles_los}, where the velocities of the Si peaks drop as the polar angle increases away from the original He ignition point.
At any given polar angle, these peaks are generally found at higher velocities for progenitors with more massive cores.
Additionally, the outer Si peak velocity, along with that of Ca, increases with shell thickness and mass.

Extra material and burning can be seen within a few degrees along the positive and negative symmetry axis in Figure \ref{fig:2dejecta}.
This is mostly due to the colliding He shell detonation with the symmetry axis and is considered to be a likely exaggerated but unavoidable effect in 2-dimensional simulations.
3D double detonation simulations from \cite{gronow_sne_2020} show similar but less pronounced radial protrusions along the ignition axis, but it is unclear whether this is attributable to the He detonation or the outgoing core detonation shock.

We note that the outermost edge ejecta of the thickest shell in Figure \ref{fig:1dprofiles_samemass} does not quite return to the initial He shell composition as in all the other presented ejecta.
This is related to the artifacts apparent for the same run in Figure \ref{fig:2dejecta} and is a result of poor shell particle sampling in our thick shell runs.
Since individual particles trace relatively low amounts of material, just $5.4 \times 10^{-6}$ M$_{\odot}$ in our thickest shell model, these areas of poor sampling represent areas of low mass despite a wide velocity extent.
Regardless, future studies should increase the number of particles initialized in the shell to maintain the satisfactory particle-to-mass ratio as in our thin-shell cases.

\subsection{Reprocessing of shell material} \label{sec:reprocessing}

For some of the very densest shell material, there are two distinct burning stages.
The first burning stage is the typical shell detonation previously described.
The second burning stage occurs when the shock from the core detonation travels outward through the He shell ashes.
For most of the shell material, this shock triggers little to no burning.
The densest material at the bottom of the shell, however, is reprocessed by this shock and burning is extended.

\begin{figure}
\centering
\includegraphics{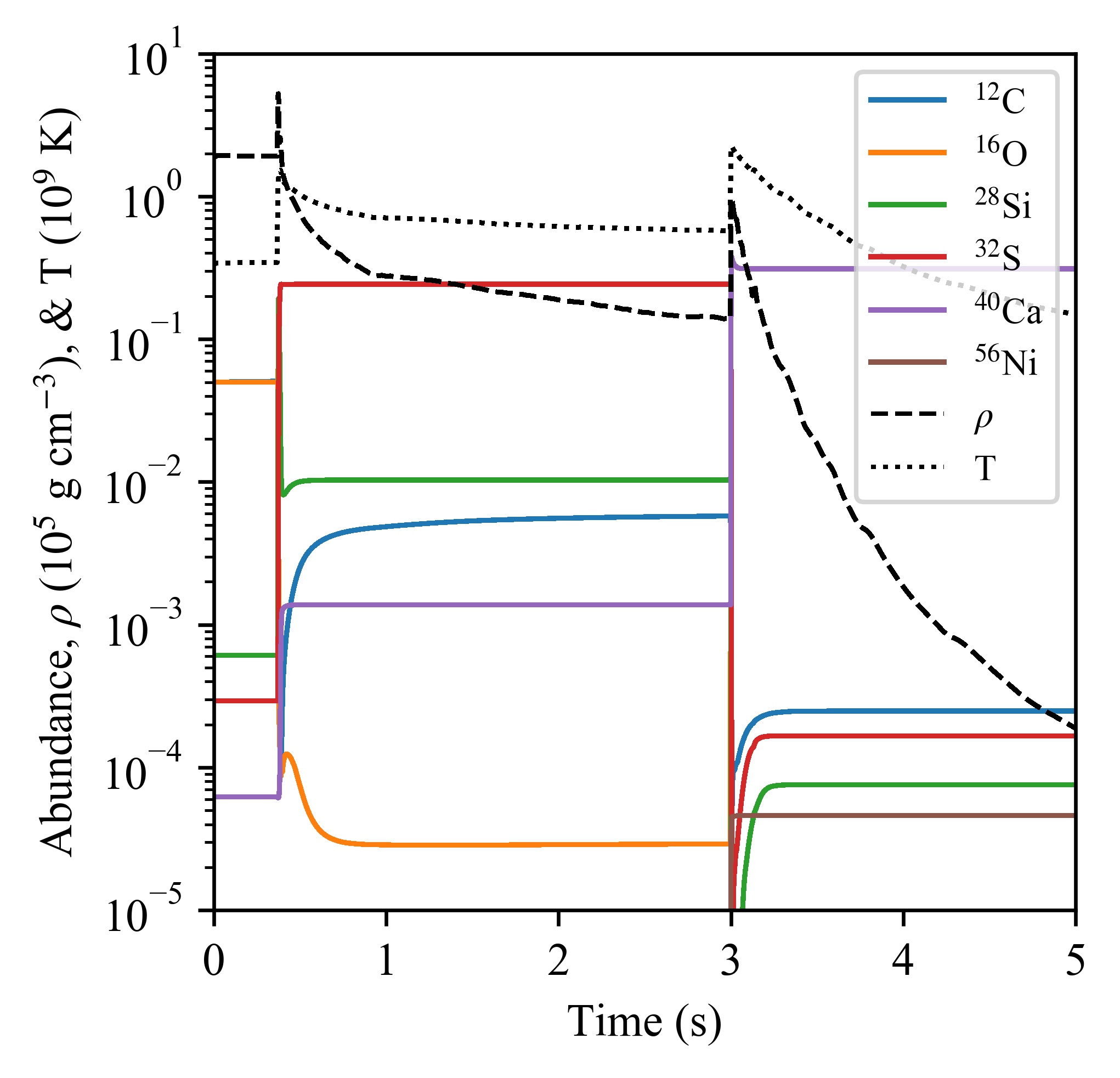}
\caption{Abundance (mass fraction) and state evolution of a single tracer that began the simulation in the shell near the core-shell interface in the $1.10 \, M_\odot$, $0.08 \, M_\odot$ run.
Two distinct burning stages occur at 0.4 and 3.0 s, corresponding with the He detonation and the remnant shock from the core detonation respectively.}
\label{fig:shell_particle}
\end{figure}

This two-phase burning process can be observed in Figure \ref{fig:shell_particle} which presents the abundance evolution of an individual tracer particle that began just above the core-shell interface.
The interface is at a density of $2\times 10^5$ g cm$^{-3}$, while this tracer started at a density of $1.93\times 10^{5}$ g cm$^{-3}$ and at a polar angle of 58$^{\circ}$.
The two burning stages occur at 0.4 and 3.0 s and correspond with the He detonation and the remnant shock from the core detonation respectively.
In this case, shell material that originally burned to mostly S winds up as predominantly Ca and unburned He with significant traces (abundance $>$ 0.01)  of radioactive Ti and Cr as well.
The shock from the core detonation is not a propagating detonation at this time, but is nevertheless sufficiently strong to cause further nucleosynthetic processing of the shell ashes.

The tracer history shown in Figure \ref{fig:shell_particle} was chosen as one for which the effects of the reprocessing are particularly profound, but often the secondary shock has little to no effect on the final abundance of the shell material.
The degree of reprocessing is dependent on material density, location, and core mass.
Significant reprocessing is only observed for the densest material close to the interface and falls away quickly with radius.
Additionally, we find a polar angle dependence of this effect which is similar to other aspects of the ejecta.
Specifically, stronger post-secondary detonation shell material burning occurs the closer the material is to the ignition pole.

An example of this effect can be seen the $M_{\rm tot} = 1.10 \, M_\odot$, $M_{\rm sh} = 0.011 \, M_\odot$ case in Figure \ref{fig:2dejecta}.
There is a compact region of elevated $^{48}$Cr abundance in the northern region of the shell ashes.
Progenitors with more massive cores have more significant reprocessing due to the stronger corresponding outgoing shock.
We attribute the dependence on polar angle to both the degree of alignment of the outgoing shock with the radial stellar density gradient, causing more shock strengthening \citep{Miles_etal_2019}, as well as the more planar nature of the secondary detonation shock near the polar region (see panels 4 and 5 in Figure \ref{fig:walkthrough}).
Mixing between shell ashes and outer core material, which persists the longest near the ignition pole, may also enhance this reprocessing effect.

\subsection{Comparison to our previous work}
\label{subsec:t19_comp}
For comparison to the single simulation from \cite{Townsley_2019a}, we examine the $1.02 \, M_\odot$ progenitor from our work that used our standard He shell abundances.
The progenitors are identical between the two simulations, but \cite{Townsley_2019a} did not implement a burning limiter and used tracer particle sampling that was more coarse.
One notable difference in the outcomes of these two simulations is a modest difference in total $^{56}$Ni yield.
Our simulation generates $0.065 \, M_\odot$, about 10\%, less $^{56}$Ni than that of \cite{Townsley_2019a}.

We believe this difference may be attributed to the use of a limiter in our study.
Other yields are slightly different but are still relatively consistent compared to the differences between simulations in this work.
Based on the work by \citet{Miles_etal_2019}, we consider the differences between this work and that of \cite{Townsley_2019a} to be consistent with those expected from the differing treatments of the under-resolved detonations.
Use of non-limited burning is expected to slightly overpredict the completeness of burning, while use of a limiter slightly underpredicts the same (see Figure 1 in \citealt{Shen_2018a} for how the choice of limiter strength in a similarly-implemented burning limiter affects the yields).
We consider the yields using a limiter to be closer to the correct yields and sufficient for current purposes.
More accurate yields would require more advanced techniques for treating the unresolved, curved detonation front in the presence of density gradients as discussed in \citet{Miles_etal_2019}.
The choice of progenitor is clearly the dominant factor in influencing the final abundances among the cases we study here.

\subsection{Comparison to other work}

Few studies, especially those in multiple dimensions, reach comparably low shell masses of our progenitors due to less-complete nuclear networks preventing the He shell detonation and as such, direct comparisons are limited.
The bulk yields in our runs of $1.0 \, M_\odot$ and above progenitors are roughly consistent with previous 1-dimensional simulations of detonations in sub-Chandrasekhar WDs of similar masses, both with He shells \citep{woosley_2011,Polin_2019} and without \citep{sim_2010,Shen_2018a,Miles_etal_2019,kushnir_etal_2020}.
The $^{56}$Ni yields for the $0.85 \, M_\odot$ runs in this study disagree with results of similar progenitors in some previous studies, producing about twice as much as that in \citet{Polin_2019} and \citet{sim_2010}.
They are, however, more consistent with yields from \cite{Shen_2018a} and \citet{Miles_etal_2019} that simulated core ignitions in bare WDs.
Bulk $^{56}$Ni yields from our $1.0 \, M_\odot$ models are roughly consistent with the 3-dimensional study of \citet{gronow_sne_2020}, but are notably higher than that from \citet{garcia_senz_2018}.
(See \citealt{Shen_2018a} for a detailed comparison of yields from sub-Chandrasekhar simulations by various authors, including some that disagree.)

With regards to HME produced in the shell, our models produce much less than that from \cite{Polin_2019} for the very thinnest shell models.
For example, the $1.00 \, M_\odot$, $0.02 \, M_\odot$ shell model from \cite{Polin_2019} produced over three orders of magnitudes more $^{56}$Ni in the shell than comparable models from this work.
Results for the thick shell yields are more consistent between our work and \cite{Polin_2019}, in addition to those from \cite{woosley_2011}.
The thickest shell model from this work produces over a magnitude more $^{56}$Ni in the shell than comparable, albeit rotating, progenitors in \citet{garcia_senz_2018}.

Again, direct comparisons of ejecta profiles are difficult due to a lack of variety of profiles shown in works from previous multi-dimensional and (comparably) thin-shell double detonation studies.
We compare our thickest shell model ($0.9 \, M_\odot$ core, $0.1 \, M_\odot$ shell) to model 9C ($0.9 \, M_\odot$ core, $0.12 \, M_\odot$ shell) from \cite{woosley_2011}.
In our thick shell model, $^{28}$Si can be found starting at around 10,000 km s$^{-1}$ at the southern pole.
The extent of the $^{28}$Si ejecta layers reach upwards of 15-18,000 km s$^{-1}$, depending on polar angle.
In model 9C from \cite{woosley_2011}, $^{28}$Si is prevalent (abundance $>$ 0.01) between 7,000 and 15,000 km s$^{-1}$.
Additionally, $^{56}$Ni can be found in our thick shell model ejecta between 11,000 and 26,000 km s$^{-1}$, again dependent on polar angle.
In the \cite{woosley_2011} 1-dimensional counterpart, $^{56}$Ni is prevalent between 10,000 and 23,000 km s$^{-1}$.
As discussed in section \ref{sec:profiles}, there are also some differences attributable to mixing at the core-shell boundary.

\section{Discussion and Conclusions}
\label{sec:conclusions}

\subsection{High-velocity material}
Regardless of total progenitor mass, very little HMEs or radioactive material are generated in the shell of our thin-shell models.
Rather, the shell ashes are heavily dominated by IMEs, especially Si and Ca. 
This is consistent with the high velocity features observed in SNe Ia \citep{childress_2014,maguire__2014,silverman_2015} given that the material generated in the shell is found at the highest velocities.
The lack of HMEs generated in the shell region of our models bolsters the potential for the double detonation scenario contributing to observed SNe Ia.
While radiative transfer calculations must be completed to directly compare to observation, the yields and profiles of our thin-shell models indicate a range of viable progenitors provided a sufficiently-low shell base density.

\subsection{Si velocities and line of sight}
If SNe Ia in nature produce asymmetric ejecta as in our models, line of sight to an observed event would certainly have an effect on the observables as in the singular case shown in Figures 3 and 4 of \citet{Townsley_2019a}.

\cite{zhang_distribution_2020} examined Si {\sc ii} absorption velocities among observed SNe Ia and showed that their distribution is bimodal.
This may be due to separate populations (symmetric and asymmetric) in SNe Ia and/or the nature of asymmetric ejecta itself. 
\cite{zhang_distribution_2020} conducted statistical simulations using line of sight-dependent Si {\sc ii} velocities parameterized by the velocity range and critical polar angle beyond which the velocity remains relatively constant.
This parametrization is able to fit the line velocity results from our previous work in \cite{Townsley_2019a} (see Figure 4 in \citealt{zhang_distribution_2020}).
It was found that the observed distribution cannot be explained with only the distribution implied by the $1.0 \, M_\odot$ progenitor and the assumption of a random line-of-sight.
Either a separate, less-asymmetric, population or an angular distribution with high Si velocities more constrained to the polar region than found by \citet{Townsley_2019a} was required.

The asymmetric ejecta in this work closely resembles that of the single run in \cite{Townsley_2019a}.
Thus, we expect the Si {\sc ii} line velocities from most of our models to resemble the line of sight-dependent model used by \cite{zhang_distribution_2020} to represent those results.
When comparing the Si {\sc ii} $\lambda$  6355 velocities from \cite{Townsley_2019a} to the inner Si peak ejection velocities, they are consistent within a couple thousands of km s$^{-1}$ across viewing angles.
We expect that the spectral line velocity will be connected to the velocity where the outer core-produced Si peaks, for any given viewing angle.
From the comparison of equatorial profiles of models at a single mass with different He shell masses in Figure \ref{fig:1dprofiles_samemass}, the extent of the core-produced Si does not depend strongly on the He shell thickness.
However, as seen in both the 2-dimensional ejecta (Figure \ref{fig:2dejecta}) and the comparison among equatorial profiles with different WD masses (Figure \ref{fig:1dprofiles_samedens}), the nature of the asymmetry is sensitive to the WD mass.
The additional dependence on mass presents the interesting possibility that the overall expected Si velocity distribution may not be just that seen from the $1.0 \, M_\odot$ progenitor.
Precise analysis of the absorption lines demands radiative transfer calculations, planned for future follow-up work.

\citet{li_can_2021} also acknowledged two populations of SNe Ia, within a sample of 16, delineated by their Si {\sc ii} velocities and showed that the high Si velocity ($>$ 11,000~km~s$^{-1}$) SNe Ia are exclusively redshifted in their nebular phase, while so-called normal Si velocity SNe Ia are predominantly blueshifted (see Figure 7 of \citealt{li_can_2021}).
They claim that this is reconcilable with the sub-Chandrasekhar He-shell model in that high velocity Si {\sc ii} observations originate from lines of sight pointed towards the shell ignition point (north), and lower velocity observations from the opposite (south), which is consistent with our asymmetric distribution of shell ashes.
Since our ejecta has higher core ash densities in the southern hemisphere than in the northern, this may also lead to a similar line of sight-nebular velocity shift relationship as seen in \citet{li_can_2021}.
Additionally, there appears to be a strong correlation between host galaxy mass and nebular velocity shift \citep{li_can_2021}. 
This notable trend demands future simulations that explore the double detonation model with varying progenitor metallicities.

The outer, high-velocity portion of the ejecta presents even more asymmetry than that in which the Si {\sc ii} lines are primarily formed.
These present possible opportunities for spectral signatures indicative of specific aspects of the He shell detonation.
One particularly interesting aspect of the asymmetric ejecta is the line-of-sight dependence caused by the reprocessing described in Section \ref{sec:reprocessing}.
This leads to the prevalence of some material in the northern hemisphere, like Ca, that would not otherwise be produced in some runs.
If the double detonation scenario contributes to a number of observed SNe Ia, it is possible that this reprocessing may introduce a line-of-sight effect on the spectra and/or photometry that is unique to the double-detonation mechanism.
This nature of this mechanism encourages further analysis and consideration across a wider range of potential progenitor models with various shell characteristics.

In addition to a general line-of-sight effect, the asymmetry in the ejecta presented in this work may result in detectable levels of polarization. 
\citet{bulla_2016} showed that asymmetrical ejecta, including a double detonation sub-Chandrasekhar model from \citet{Fink_10}, can reproduce some subtle polarization features observed in SNe Ia. 
It is challenging to predict exactly what degree of polarization will result from our new ejecta, but we expect future calculations to show that the polarization of our ejecta will generally resemble the modest polarization shown in \citet{bulla_2016}, given the similarities between the core ejecta of \citet{Fink_10} and our ejecta.
Deviations from the double detonation polarization presented in \citet{bulla_2016} will not be surprising, however, especially features that may be attributed to high velocity or shell ash material.

\subsection{Conclusions}
We have simulated double detonations in eleven unique progenitors across a range of core masses and shell thickness, focusing on the thin-shell candidate model.
We have presented the results from these simulations in bulks yields and ejecta profiles.
In summary, our main conclusions are:
\begin{itemize}

    \item Yields of our thin-shell progenitors are similar to \cite{Townsley_2019a}, indicating that spectroscopic features of these progenitors are likely to be consistent with observed SNe Ia.
    Cases with different WD masses but equal He shell base densities produce a similar degree of HMEs in the shell.

    \item HME yields, including $^{56}$Ni, increase with progenitor mass and cover the range of expectations for normal SNe Ia.

    \item Shell yields are very sensitive to the choice of shell base parameters, with shells with a base density between 3 and $6 \times 10^{5}$ g cm$^{-3}$ starting to produce significant amounts of elements that may lead to spectral abnormalities.
    This will constrain the allowed shell thickness.

    \item The core is ignited, in at least some cases, some distance away from the symmetry axis, confirming 1-dimensional work \citep{Kushnir_etal_2012,Shen_2014a}, an aspect that requires further analysis in multi-dimensional studies.

    \item Ejecta from our simulations are modestly but significantly asymmetrical, leading to a  line-of-sight dependence of the ejecta velocity extent of certain elements, including Si and Ca.
    
    \item Non-trivial burning may occur in the shell after the shell detonation is complete by the shock from the core detonation passing out into the partially-burned shell.
    Due to the varying angle of the outgoing shock, this results in further asymmetrization of the ejection.
    
    \item Comparison of ejecta profiles to previous 1-dimensional studies reveals differences near the core-shell boundary that are due to shear mixing induced by the He detonation.
    
\end{itemize}

Our simulated double detonations show that sub-Chandrasekhar WDs with a range of He shell parameters and WD masses are viable candidates for some portion of observed SNe Ia judging by the simulations yields and ejecta structure.
To further verify the likelihood of the contribution of double detonations on the population of observed SNe Ia, radiative transfer calculations must be performed using these ejecta in order to determine how these models truly compare to reality.


\acknowledgements

We thank the referee for their helpful comments.
This work was supported by the NASA Astrophysics Theory Program (NNX17AG28G).
Computations were performed on NASA's Pleiades and institutional resources at The University of Alabama.
S.J.B. acknowledges support from NASA grant HST-AR-16156.
We thank Carla Fr\"ohlich for supporting B.J.M.'s contribution to this work.

Portions of this work were supported by the United States Department of Energy, under an Early Career Award (Grant No. SC0010263), by the Office of Science, Office of Nuclear Physics award DE-FG02-02ER41216, and by the Research Corporation for Science Advancement under a Cottrell Scholar Award.

\software{\texttt{FLASH} (\citealt{Fryxelletal2000,Dubeyetal2009,Dubeyetal2013,Dubeyetal2014}, \url{flash.uchicago.edu}),
\texttt{MESA} (\citealt{paxton_2011,paxton_2013,paxton_2015,paxton_2018}, \url{mesa.sourceforge.net}),
\texttt{yt} (\url{yt-project.org}})

\bibliography{references}{}
\bibliographystyle{aasjournal}

\appendix

Extended yields at 100 s for species that have a core or shell abundance above $10^{-8}$ in any run at 100 s are given in Tables \ref{table:fullyields_a}--\ref{table:fullyields_d}.
Asymptotic yields at 1 Gyr are given in Tables \ref{table:asymptotic_yields_a}--\ref{table:asymptotic_yields_d}.
The only isotope that is not stable or effectively stable on this timescale is $^{40}$K.

\setlength{\tabcolsep}{6pt}

\centering
\begin{longtable}{c|cc|cc|cc|cc}
\caption{More Yields A}\\
\multicolumn{1}{c|}{} & \multicolumn{2}{c|}{d2e5\_m085} & \multicolumn{2}{c|}{d2e5\_m090} & \multicolumn{2}{c|}{d2e5\_m100} & \multicolumn{2}{c}{d2e5\_m110} \\
 &  shell &  total &  shell &  total &  shell & total & shell & total \\
 &($M_\odot$) & ($M_\odot$) & ($M_\odot$) & ($M_\odot$) & ($M_\odot$) & ($M_\odot$) & ($M_\odot$) & ($M_\odot$) \\ \hline
All & 0.033 & 0.85 & 0.025 & 0.90 & 0.016 & 1.00 & 8.4$\times 10^{-3}$ & 1.10 \\ \hline
\endfirsthead
\multicolumn{1}{c|}{} & \multicolumn{2}{c|}{d2e5\_m085} & \multicolumn{2}{c|}{d2e5\_m090} & \multicolumn{2}{c|}{d2e5\_m100} & \multicolumn{2}{c}{d2e5\_m110} \\
 &  shell &  total &  shell &  total &  shell & total & shell & total \\ \hline
\endhead
$^{4}$He & 0.023 & 0.023 & 0.018 & 0.018 & 0.012 & 0.016 & 6.2$\times 10^{-3}$ & 0.015 \\ 
$^{12}$C & 1.9$\times 10^{-4}$ & 0.014 & 1.8$\times 10^{-4}$ & 0.011 & 1.3$\times 10^{-4}$ & 4.1$\times 10^{-3}$ & 9.7$\times 10^{-5}$ & 1.9$\times 10^{-3}$ \\ 
$^{13}$C & 4.1$\times 10^{-11}$ & 8.9$\times 10^{-8}$ & 1.4$\times 10^{-10}$ & 8.1$\times 10^{-8}$ & 1.6$\times 10^{-10}$ & 3.5$\times 10^{-8}$ & 2.1$\times 10^{-10}$ & 1.6$\times 10^{-8}$ \\ 
$^{13}$N & 1.5$\times 10^{-9}$ & 1.5$\times 10^{-9}$ & 1.6$\times 10^{-9}$ & 1.6$\times 10^{-9}$ & 1.3$\times 10^{-9}$ & 1.3$\times 10^{-9}$ & 1.7$\times 10^{-9}$ & 1.8$\times 10^{-9}$ \\ 
$^{14}$N & 2.2$\times 10^{-6}$ & 1.0$\times 10^{-5}$ & 3.1$\times 10^{-6}$ & 9.1$\times 10^{-6}$ & 2.3$\times 10^{-6}$ & 5.3$\times 10^{-6}$ & 3.4$\times 10^{-6}$ & 4.6$\times 10^{-6}$ \\ 
$^{15}$N & 1.3$\times 10^{-7}$ & 1.5$\times 10^{-7}$ & 4.1$\times 10^{-7}$ & 4.3$\times 10^{-7}$ & 4.8$\times 10^{-7}$ & 4.9$\times 10^{-7}$ & 5.2$\times 10^{-7}$ & 5.2$\times 10^{-7}$ \\ 
$^{15}$O & 8.9$\times 10^{-7}$ & 8.9$\times 10^{-7}$ & 7.9$\times 10^{-7}$ & 7.9$\times 10^{-7}$ & 6.4$\times 10^{-7}$ & 6.4$\times 10^{-7}$ & 7.0$\times 10^{-7}$ & 7.0$\times 10^{-7}$ \\ 
$^{16}$O & 8.7$\times 10^{-5}$ & 0.170 & 9.9$\times 10^{-5}$ & 0.137 & 8.3$\times 10^{-5}$ & 0.079 & 7.5$\times 10^{-5}$ & 0.041 \\
\end{longtable}

\centering
\begin{longtable}{c|cc|cc}
\caption{More Yields B}\\
\multicolumn{1}{c|}{} & \multicolumn{2}{c|}{d2e5\_m102} & \multicolumn{2}{c}{d2e5\_m102\_lowO} \\
 &  shell &  total &  shell &  total \\
 &($M_\odot$) & ($M_\odot$) & ($M_\odot$) & ($M_\odot$) \\ \hline
All & 0.021 & 1.02 & 0.021 & 1.02 \\ \hline
\endfirsthead
\multicolumn{1}{c|}{} & \multicolumn{2}{c|}{d2e5\_m102} & \multicolumn{2}{c}{d2e5\_m102\_lowO} \\
 &  shell &  total &  shell &  total \\ \hline
\endhead
$^{4}$He & 0.015 & 0.020 & 0.016 & 0.022 \\ 
$^{12}$C & 1.4$\times 10^{-4}$ & 2.9$\times 10^{-3}$ & 2.7$\times 10^{-4}$ & 4.2$\times 10^{-3}$ \\ 
$^{13}$C & 5.1$\times 10^{-11}$ & 2.6$\times 10^{-8}$ & 1.1$\times 10^{-10}$ & 2.5$\times 10^{-8}$ \\ 
$^{13}$N & 1.3$\times 10^{-9}$ & 1.3$\times 10^{-9}$ & 3.4$\times 10^{-9}$ & 3.5$\times 10^{-9}$ \\ 
$^{14}$N & 2.6$\times 10^{-6}$ & 4.7$\times 10^{-6}$ & 5.5$\times 10^{-6}$ & 8.0$\times 10^{-6}$ \\ 
$^{15}$N & 1.5$\times 10^{-7}$ & 1.6$\times 10^{-7}$ & 3.0$\times 10^{-7}$ & 3.0$\times 10^{-7}$ \\ 
$^{15}$O & 7.4$\times 10^{-7}$ & 7.5$\times 10^{-7}$ & 1.7$\times 10^{-6}$ & 1.7$\times 10^{-6}$ \\ 
$^{16}$O & 7.3$\times 10^{-5}$ & 0.070 & 7.5$\times 10^{-5}$ & 0.075 \\
\end{longtable}

\centering
\begin{longtable}{c|cc|cc|cc}
\caption{More Yields C}\\
\multicolumn{1}{c|}{} & \multicolumn{2}{c|}{d3e5\_m085} & \multicolumn{2}{c|}{d3e5\_m100} & \multicolumn{2}{c}{d3e5\_m110} \\
 &  shell &  total &  shell &  total &  shell & total \\
 &($M_\odot$) & ($M_\odot$) & ($M_\odot$) & ($M_\odot$) & ($M_\odot$) & ($M_\odot$)  \\ \hline
All & 0.049 & 0.85 & 0.021 & 1.00 & 0.011 & 1.10 \\ \hline
\endfirsthead
\multicolumn{1}{c|}{} & \multicolumn{2}{c|}{d3e5\_m085} & \multicolumn{2}{c|}{d3e5\_m100} & \multicolumn{2}{c}{d3e5\_m110} \\
 &  shell &  total &  shell &  total &  shell & total \\ \hline
\endhead
$^{4}$He & 0.030 & 0.030 & 0.014 & 0.019 & 7.3$\times 10^{-3}$ & 0.017 \\ 
$^{12}$C & 1.3$\times 10^{-4}$ & 0.011 & 5.9$\times 10^{-5}$ & 3.2$\times 10^{-3}$ & 3.2$\times 10^{-5}$ & 1.4$\times 10^{-3}$ \\ 
$^{13}$C & 4.8$\times 10^{-11}$ & 8.7$\times 10^{-8}$ & 2.6$\times 10^{-11}$ & 3.1$\times 10^{-8}$ & 1.9$\times 10^{-11}$ & 1.7$\times 10^{-8}$ \\ 
$^{13}$N & 3.9$\times 10^{-10}$ & 4.0$\times 10^{-10}$ & 2.1$\times 10^{-10}$ & 2.3$\times 10^{-10}$ & 1.6$\times 10^{-10}$ & 1.8$\times 10^{-10}$ \\ 
$^{14}$N & 3.1$\times 10^{-7}$ & 8.7$\times 10^{-6}$ & 2.0$\times 10^{-7}$ & 2.2$\times 10^{-6}$ & 9.9$\times 10^{-8}$ & 9.5$\times 10^{-7}$ \\ 
$^{15}$N & 7.4$\times 10^{-8}$ & 9.0$\times 10^{-8}$ & 5.2$\times 10^{-8}$ & 5.6$\times 10^{-8}$ & 3.5$\times 10^{-8}$ & 3.6$\times 10^{-8}$ \\ 
$^{15}$O & 9.8$\times 10^{-8}$ & 1.0$\times 10^{-7}$ & 6.9$\times 10^{-8}$ & 6.9$\times 10^{-8}$ & 4.5$\times 10^{-8}$ & 4.6$\times 10^{-8}$ \\ 
$^{16}$O & 2.5$\times 10^{-5}$ & 0.142 & 1.4$\times 10^{-5}$ & 0.074 & 1.0$\times 10^{-5}$ & 0.039 \\
\end{longtable}

\centering
\begin{longtable}{c|cc|cc}
\caption{More Yields D}\\
\multicolumn{1}{c|}{} & \multicolumn{2}{c|}{d6e5\_m100} & \multicolumn{2}{c}{d14e5\_m100} \\
 &  shell &  total &  shell &  total \\
 &($M_\odot$) & ($M_\odot$) & ($M_\odot$) & ($M_\odot$) \\ \hline
All & 0.042 & 1.00 & 0.100 & 1.00 \\ \hline
\endfirsthead
\multicolumn{1}{c|}{} & \multicolumn{2}{c|}{d6e5\_m100} & \multicolumn{2}{c}{d14e5\_m100} \\
 &  shell &  total &  shell &  total \\ \hline
\endhead
$^{4}$He & 0.020 & 0.025 & 0.030 & 0.036 \\ 
$^{12}$C & 1.1$\times 10^{-5}$ & 2.4$\times 10^{-3}$ & 1.2$\times 10^{-6}$ & 5.9$\times 10^{-4}$ \\ 
$^{13}$C & 7.6$\times 10^{-10}$ & 2.0$\times 10^{-8}$ & 2.3$\times 10^{-9}$ & 2.3$\times 10^{-9}$ \\ 
$^{13}$N & 6.4$\times 10^{-9}$ & 6.4$\times 10^{-9}$ & 1.9$\times 10^{-8}$ & 1.9$\times 10^{-8}$ \\ 
$^{14}$N & 2.2$\times 10^{-10}$ & 1.7$\times 10^{-6}$ & 5.3$\times 10^{-10}$ & 1.3$\times 10^{-8}$ \\ 
$^{15}$N & 1.1$\times 10^{-9}$ & 3.1$\times 10^{-9}$ & 3.7$\times 10^{-10}$ & 6.8$\times 10^{-10}$ \\ 
$^{15}$O & 1.5$\times 10^{-9}$ & 2.2$\times 10^{-9}$ & 5.2$\times 10^{-10}$ & 9.3$\times 10^{-10}$ \\ 
$^{16}$O & 5.6$\times 10^{-6}$ & 0.069 & 5.4$\times 10^{-6}$ & 0.068 \\
\end{longtable}

\centering
\begin{longtable}{c|cc|cc|cc|cc}
\caption{Asymptotic Decay Yields A}\\
\multicolumn{1}{c|}{} & \multicolumn{2}{c|}{d2e5\_m085} & \multicolumn{2}{c|}{d2e5\_m090} & \multicolumn{2}{c|}{d2e5\_m100} & \multicolumn{2}{c}{d2e5\_m110} \\
 &  shell &  total &  shell &  total &  shell & total & shell & total \\
 &($M_\odot$) & ($M_\odot$) & ($M_\odot$) & ($M_\odot$) & ($M_\odot$) & ($M_\odot$) & ($M_\odot$) & ($M_\odot$) \\ \hline
All & 0.033 & 0.85 & 0.025 & 0.90 & 0.016 & 1.00 & 8.4$\times 10^{-3}$ & 1.10 \\ \hline
\endfirsthead
\multicolumn{1}{c|}{} & \multicolumn{2}{c|}{d2e5\_m085} & \multicolumn{2}{c|}{d2e5\_m090} & \multicolumn{2}{c|}{d2e5\_m100} & \multicolumn{2}{c}{d2e5\_m110} \\
 &  shell &  total &  shell &  total &  shell & total & shell & total \\ \hline
\endhead
$^{12}$C & 1.9$\times 10^{-4}$ & 0.014 & 1.8$\times 10^{-4}$ & 0.011 & 1.3$\times 10^{-4}$ & 4.1$\times 10^{-3}$ & 9.7$\times 10^{-5}$ & 1.9$\times 10^{-3}$ \\ 
$^{13}$C & 1.5$\times 10^{-9}$ & 9.0$\times 10^{-8}$ & 1.7$\times 10^{-9}$ & 8.2$\times 10^{-8}$ & 1.5$\times 10^{-9}$ & 3.6$\times 10^{-8}$ & 1.9$\times 10^{-9}$ & 1.8$\times 10^{-8}$ \\ 
$^{14}$N & 2.2$\times 10^{-6}$ & 1.0$\times 10^{-5}$ & 3.1$\times 10^{-6}$ & 9.1$\times 10^{-6}$ & 2.3$\times 10^{-6}$ & 5.3$\times 10^{-6}$ & 3.4$\times 10^{-6}$ & 4.6$\times 10^{-6}$ \\ 
$^{15}$N & 1.0$\times 10^{-6}$ & 1.0$\times 10^{-6}$ & 1.2$\times 10^{-6}$ & 1.2$\times 10^{-6}$ & 1.1$\times 10^{-6}$ & 1.1$\times 10^{-6}$ & 1.2$\times 10^{-6}$ & 1.2$\times 10^{-6}$ \\ 
$^{16}$O & 8.7$\times 10^{-5}$ & 0.170 & 9.9$\times 10^{-5}$ & 0.137 & 8.3$\times 10^{-5}$ & 0.079 & 7.5$\times 10^{-5}$ & 0.041 \\ 
$^{17}$O & 4.4$\times 10^{-10}$ & 5.3$\times 10^{-6}$ & 5.5$\times 10^{-10}$ & 4.2$\times 10^{-6}$ & 3.7$\times 10^{-10}$ & 1.9$\times 10^{-6}$ & 7.8$\times 10^{-10}$ & 7.2$\times 10^{-7}$ \\ 
$^{18}$O & 1.6$\times 10^{-6}$ & 1.7$\times 10^{-6}$ & 2.4$\times 10^{-6}$ & 2.5$\times 10^{-6}$ & 1.9$\times 10^{-6}$ & 1.9$\times 10^{-6}$ & 2.9$\times 10^{-6}$ & 2.9$\times 10^{-6}$ \\ 
$^{19}$F & 9.3$\times 10^{-7}$ & 9.4$\times 10^{-7}$ & 1.2$\times 10^{-6}$ & 1.2$\times 10^{-6}$ & 1.1$\times 10^{-6}$ & 1.1$\times 10^{-6}$ & 1.1$\times 10^{-6}$ & 1.1$\times 10^{-6}$ \\
\end{longtable}

\centering
\begin{longtable}{c|cc|cc}
\caption{Asymptotic Decay Yields B}\\
\multicolumn{1}{c|}{} & \multicolumn{2}{c|}{d2e5\_m102} & \multicolumn{2}{c}{d2e5\_m102\_lowO} \\
 &  shell &  total &  shell &  total \\
 &($M_\odot$) & ($M_\odot$) & ($M_\odot$) & ($M_\odot$) \\ \hline
All & 0.021 & 1.02 & 0.021 & 1.02 \\ \hline
\endfirsthead
\multicolumn{1}{c|}{} & \multicolumn{2}{c|}{d2e5\_m102} & \multicolumn{2}{c}{d2e5\_m102\_lowO} \\
 &  shell &  total &  shell &  total \\ \hline
\endhead
$^{12}$C & 1.4$\times 10^{-4}$ & 2.9$\times 10^{-3}$ & 2.7$\times 10^{-4}$ & 4.2$\times 10^{-3}$ \\ 
$^{13}$C & 1.4$\times 10^{-9}$ & 2.8$\times 10^{-8}$ & 3.6$\times 10^{-9}$ & 2.8$\times 10^{-8}$ \\ 
$^{14}$N & 2.6$\times 10^{-6}$ & 4.7$\times 10^{-6}$ & 5.5$\times 10^{-6}$ & 8.0$\times 10^{-6}$ \\ 
$^{15}$N & 9.0$\times 10^{-7}$ & 9.0$\times 10^{-7}$ & 2.0$\times 10^{-6}$ & 2.0$\times 10^{-6}$ \\ 
$^{16}$O & 7.3$\times 10^{-5}$ & 0.070 & 7.5$\times 10^{-5}$ & 0.075 \\ 
$^{17}$O & 4.7$\times 10^{-10}$ & 1.3$\times 10^{-6}$ & 1.2$\times 10^{-9}$ & 1.6$\times 10^{-6}$ \\ 
$^{18}$O & 1.7$\times 10^{-6}$ & 1.8$\times 10^{-6}$ & 3.8$\times 10^{-6}$ & 3.8$\times 10^{-6}$ \\ 
$^{19}$F & 8.2$\times 10^{-7}$ & 8.2$\times 10^{-7}$ & 1.9$\times 10^{-6}$ & 1.9$\times 10^{-6}$ \\
\end{longtable}

\centering
\begin{longtable}{c|cc|cc|cc}
\caption{Asymptotic Decay Yields C}\\
\multicolumn{1}{c|}{} & \multicolumn{2}{c|}{d3e5\_m085} & \multicolumn{2}{c|}{d3e5\_m100} & \multicolumn{2}{c}{d3e5\_m110} \\
 &  shell &  total &  shell &  total &  shell & total \\
 &($M_\odot$) & ($M_\odot$) & ($M_\odot$) & ($M_\odot$) & ($M_\odot$) & ($M_\odot$)  \\ \hline
All & 0.049 & 0.85 & 0.021 & 1.00 & 0.011 & 1.10 \\ \hline
\endfirsthead
\multicolumn{1}{c|}{} & \multicolumn{2}{c|}{d3e5\_m085} & \multicolumn{2}{c|}{d3e5\_m100} & \multicolumn{2}{c}{d3e5\_m110} \\
 &  shell &  total &  shell &  total &  shell & total \\ \hline
\endhead
$^{12}$C & 1.3$\times 10^{-4}$ & 0.011 & 5.9$\times 10^{-5}$ & 3.2$\times 10^{-3}$ & 3.2$\times 10^{-5}$ & 1.4$\times 10^{-3}$ \\ 
$^{13}$C & 4.4$\times 10^{-10}$ & 8.8$\times 10^{-8}$ & 2.4$\times 10^{-10}$ & 3.1$\times 10^{-8}$ & 1.8$\times 10^{-10}$ & 1.7$\times 10^{-8}$ \\ 
$^{14}$N & 3.1$\times 10^{-7}$ & 8.7$\times 10^{-6}$ & 2.0$\times 10^{-7}$ & 2.2$\times 10^{-6}$ & 9.9$\times 10^{-8}$ & 9.5$\times 10^{-7}$ \\ 
$^{15}$N & 1.7$\times 10^{-7}$ & 1.9$\times 10^{-7}$ & 1.2$\times 10^{-7}$ & 1.3$\times 10^{-7}$ & 8.0$\times 10^{-8}$ & 8.2$\times 10^{-8}$ \\ 
$^{16}$O & 2.5$\times 10^{-5}$ & 0.142 & 1.4$\times 10^{-5}$ & 0.074 & 1.0$\times 10^{-5}$ & 0.039 \\ 
$^{17}$O & 2.0$\times 10^{-10}$ & 5.0$\times 10^{-6}$ & 8.8$\times 10^{-11}$ & 1.1$\times 10^{-6}$ & 4.4$\times 10^{-11}$ & 4.5$\times 10^{-7}$ \\ 
$^{18}$O & 1.4$\times 10^{-7}$ & 2.0$\times 10^{-7}$ & 1.1$\times 10^{-7}$ & 1.2$\times 10^{-7}$ & 6.0$\times 10^{-8}$ & 6.3$\times 10^{-8}$ \\ 
$^{19}$F & 1.3$\times 10^{-7}$ & 1.4$\times 10^{-7}$ & 8.4$\times 10^{-8}$ & 8.5$\times 10^{-8}$ & 6.7$\times 10^{-8}$ & 6.7$\times 10^{-8}$ \\
\end{longtable}

\centering
\begin{longtable}{c|cc|cc}
\caption{Asymptotic Decay Yields D}\\
\multicolumn{1}{c|}{} & \multicolumn{2}{c|}{d6e5\_m100} & \multicolumn{2}{c}{d14e5\_m100} \\
 &  shell &  total &  shell &  total \\
 &($M_\odot$) & ($M_\odot$) & ($M_\odot$) & ($M_\odot$) \\ \hline
All & 0.042 & 1.00 & 0.100 & 1.00 \\ \hline
\endfirsthead
\multicolumn{1}{c|}{} & \multicolumn{2}{c|}{d6e5\_m100} & \multicolumn{2}{c}{d14e5\_m100} \\
 &  shell &  total &  shell &  total \\ \hline
\endhead
$^{12}$C & 1.1$\times 10^{-5}$ & 2.4$\times 10^{-3}$ & 1.2$\times 10^{-6}$ & 5.9$\times 10^{-4}$ \\ 
$^{13}$C & 7.2$\times 10^{-9}$ & 2.6$\times 10^{-8}$ & 2.1$\times 10^{-8}$ & 2.1$\times 10^{-8}$ \\ 
$^{14}$N & 2.2$\times 10^{-10}$ & 1.7$\times 10^{-6}$ & 5.3$\times 10^{-10}$ & 1.3$\times 10^{-8}$ \\ 
$^{15}$N & 2.6$\times 10^{-9}$ & 5.3$\times 10^{-9}$ & 8.9$\times 10^{-10}$ & 1.6$\times 10^{-9}$ \\ 
$^{16}$O & 5.6$\times 10^{-6}$ & 0.069 & 5.4$\times 10^{-6}$ & 0.068 \\ 
$^{17}$O & 7.1$\times 10^{-9}$ & 6.4$\times 10^{-7}$ & 1.3$\times 10^{-7}$ & 1.4$\times 10^{-7}$ \\ 
$^{18}$O & 7.5$\times 10^{-11}$ & 4.1$\times 10^{-9}$ & 5.4$\times 10^{-13}$ & 2.0$\times 10^{-10}$ \\ 
$^{19}$F & 1.7$\times 10^{-9}$ & 1.9$\times 10^{-9}$ & 1.1$\times 10^{-11}$ & 2.4$\times 10^{-11}$ \\
\end{longtable}



\end{document}